\documentclass{aa}

\usepackage{natbib}
\usepackage{adjustbox}
\usepackage{array}
\usepackage{booktabs}
\usepackage{chemformula}
\usepackage{dblfloatfix}
\usepackage{float}
\usepackage{graphicx}
\usepackage[colorlinks=true, linkcolor=red, citecolor=blue, urlcolor=blue]{hyperref}
\usepackage{longtable}
\usepackage{makecell}
\usepackage{multicol}
\usepackage{placeins}
\usepackage{siunitx}
\usepackage{textgreek}
\usepackage{txfonts}
\usepackage{upgreek}
\usepackage{xcolor}

\newcommand{\etal}{et\,al.}

\newcommand{\rcm}{\ensuremath{\text{cm}^{-1}}}
\newcommand{\um}{\si{\micro\meter}}

\titlerunning{Methanol isotopologues in astrophysical ices}
\authorrunning{Vyjidak \etal}

\begin{document}
\raggedbottom

   \title{Broadband infrared spectroscopy of methanol isotopologues in~pure, \ch{H2O}-rich, and \ch{CO}-rich ice analogues}

   \author{Adam Vyjidak\inst{1},
          Barbara Michela Giuliano\inst{1}, Pavol Jusko\inst{1}, Heidy M. Quitián-Lara\inst{1}, Felipe Fantuzzi\inst{2,3}, Giuseppe A. Baratta\inst{4}, Maria Elisabetta Palumbo\inst{4}, Paola Caselli\inst{1}
          }

   \institute{Max-Planck-Institut für extraterrestrische Physik,
              Gießenbachstraße 1, D-85748 Garching, Germany
         \and
             Supramolecular, Interfacial and Synthetic Chemistry, School of Natural Sciences, University of Kent, Park Wood Rd, Canterbury CT2 7NH, United Kingdom
         \and
             HUN-REN Institute for Nuclear Research (Atomki), Debrecen H-4026, Hungary
         \and
             INAF - Osservatorio Astrofisico di Catania, via Santa Sofia 78, I-95123 Catania, Italy
             }

   \date{\today}

\abstract{
\noindent Deuterium fractionation is highly efficient during the early stages of star formation, particularly in starless and prestellar cores where temperatures are low ($< \SI{10}{\kelvin}$) and molecular freeze-out onto dust grains is significant. Methanol forms early in these environments following \ch{CO} freeze-out via successive hydrogenation reactions on grain surfaces, while the production of deuterated methanol requires elevated gas-phase D/H ratios generated through dissociative recombination of deuterated \ch{H3+}. Consequently, large abundances of deuterated methanol are observed towards young stellar objects where prestellar ices have recently sublimated. Here, we present laboratory broadband infrared spectra of methanol and its isotopologues in astrophysical ice analogues, complemented by anharmonic vibrational calculations used to guide band assignments. Experiments were performed at the CASICE laboratory using a Bruker Vertex 70v spectrometer coupled to a closed-cycle helium cryostat, with isotopologue ices deposited at \SI{10}{\kelvin} under high-vacuum conditions. Infrared transmission spectra were recorded over \SIrange{6000}{30}{\rcm} (\SIrange{1.67}{333}{\um}) and compared with spectra of pure isotopologue ices. Distinctive mid-infrared band patterns are identified for each deuterated species. In particular, \ch{CH2DOH} exhibits a characteristic doublet at \SI{1293}{\rcm} and \SI{1326}{\rcm} (\SI{7.73}{\um} and \SI{7.54}{\um}), while \ch{CHD2OH} shows a similar doublet at \SI{1301}{\rcm} and \SI{1329}{\rcm} (\SI{7.69}{\um} and \SI{7.52}{\um}), both remaining largely invariant across all studied ice mixtures. These robust spectral signatures provide reliable tracers for identifying deuterated methanol in JWST observations and for constraining astrochemical gas–grain models of deuterium enrichment prior to star and planet formation.
}

\maketitle

\section{Introduction}

Methanol (\ch{CH3OH}) is a key organic molecule in the interstellar medium (ISM) and is commonly detected in dense clouds and star-forming regions \citep{Bizzocchi2014, Jorgensen2020, Punanova2022}. \ch{CH3OH} is formed primarily on the surface of dust grains via successive hydrogenation of \ch{CO} molecules \citep{Watanabe2002, Fuchs2009}. In starless and prestellar cores, catastrophic \ch{CO} freeze-out produces \ch{CO}-rich ices, enhancing surface production and enabling reactive desorption of a fraction of the freshly formed \ch{CH3OH} into the gas phase \citep{Minissale2015, Vasyunin2017}. In these regions, deuterium fractionation becomes highly effective, as low temperatures and high levels of molecular freeze-out favour the production of deuterated forms of \ch{H3+} \citep{Caselli2003, Walmsley2004}. Dissociative recombination of \ch{H2D+}, \ch{D2H+}, and \ch{D3+} increases the atomic D/H ratio in the gas phase, allowing deuterium atoms to compete with hydrogen atoms in reactions with surface \ch{CO}, producing deuterated methanol \citep{Caselli2002, Ceccarelli2014}. Methanol and its singly and doubly deuterated isotopologues have been detected in prestellar cores \citep{Bizzocchi2014, ChaconTanarro2019, Lin2023} and towards Class 0 protostars \citep{QuitianLara2024}, with triply deuterated methanol also reported in some sources \citep{Parise2004}. A significant abundance of methanol ice has been reported towards the prototypical prestellar core L1544 \citep{Goto2021}. Therefore, deuterated methanol ice is also expected to be abundant in similar environments.

Upcoming and ongoing observations from the \textit{James Webb Space Telescope} (JWST) are now enabling more detailed studies of interstellar ices, including complex organic molecules and their isotopologues \citep{Sturm2023, McClure2023}. To interpret these infrared spectra, robust laboratory reference data are required. Laboratory astrophysics experiments therefore provide key band assignments, band strengths, and temperature-dependent spectral profiles under controlled, ISM-like conditions \citep{Cuppen2024, Giuliano2014, Linnartz2015, Giuliano2016, Ligterink2018, Muller2022, Rocha2022, Dickers2025}. These benchmarks also enable identification of isotopologue-specific signatures and help to constrain the chemical composition of interstellar ices in dark and cold molecular clouds \citep{Brunken2024, Dartois2024, McClure2023, Rocha2024, Rocha2025, Spezzano2025, Tyagi2025, vanGelder2024}.

For example, laboratory investigations of solid \ch{CH2DOH}, both in pure form and in mixtures, have demonstrated the importance of measuring refractive indices, densities, band intensities, and temperature-dependent profile changes under high-vacuum conditions, supported by precise thickness determinations using laser interferometry \citep{Palumbo2015, Urso2018, Scire2019}. These studies motivate the present work, which adopts a broader scope by providing a systematic, broadband infrared characterisation of methanol and its five main isotopologues, in pure ice and within \ch{H2O}-rich and \ch{CO}-rich matrices.

In this work, we present a laboratory investigation of methanol and its isotopologues ($\ch{CH3OD}, \ch{CH2DOH}, \ch{CHD2OH}, \ch{CD3OH}, \ch{CD3OD}$) under astrophysically relevant ice conditions, supported by anharmonic vibrational calculations to assist in the assignment of fundamental, overtone, and combination bands. Throughout this paper, “methanol” refers to the fully hydrogenated isotopologue \ch{CH3OH}, whereas “deuterated methanol” denotes the isotopologues listed above. In particular, we examine pure samples and binary mixtures with \ch{H2O} or \ch{CO} deposited at \SI{10}{\kelvin} onto a silicon substrate, and follow their evolution upon heating to \SI{120}{\kelvin}. By recording infrared spectra in transmission mode over a broad spectral range of \SIrange{6000}{30}{\rcm} (\SIrange{1.67}{333}{\um}), we identify distinct band profiles, assess the influence of mixing ratios and temperature on spectral signatures, and provide benchmarks for interpreting future astronomical observations. Our results, combined with JWST data, will help constrain astrochemical models of methanol deuteration in star-forming regions and support the use of deuterated methanol as a tracer of early ice chemistry.

\section{Experimental methods}

The experiments were conducted at the Centre for Astrochemical Studies (CAS) at the Max Planck Institute for Extraterrestrial Physics, Garching, Germany, using a custom-designed apparatus. The setup integrates a closed-cycle helium cryocooler (Advanced Research Systems, ARS) with a Bruker Vertex 70v spectrometer. The experimental vacuum chamber, attached to the cryocooler and mounted directly within the spectrometer’s sample compartment, can be positioned precisely on a motorised stage.

\subsection{Experimental setup}

Infrared (IR) spectra were obtained in transmission mode over \SIrange{6000}{30}{\rcm} (\SIrange{1.67}{333}{\um}) at \SI{10}{\kelvin} using Fourier transform infrared spectroscopy (FTIR), probing ice films deposited on both sides of the silicon substrate. Each spectrum was recorded with a Bruker Vertex 70v spectrometer equipped with a deuterated triglycine sulphate (DTGS) detector, using an \SI{8}{mm} aperture and a spectral resolution of \SI{1}{\rcm}. A total of 64 scans were averaged, as increasing to 128 did not significantly improve the signal-to-noise ratio.

\begin{figure}[htbp]
    \centering
    \includegraphics[scale=0.4]{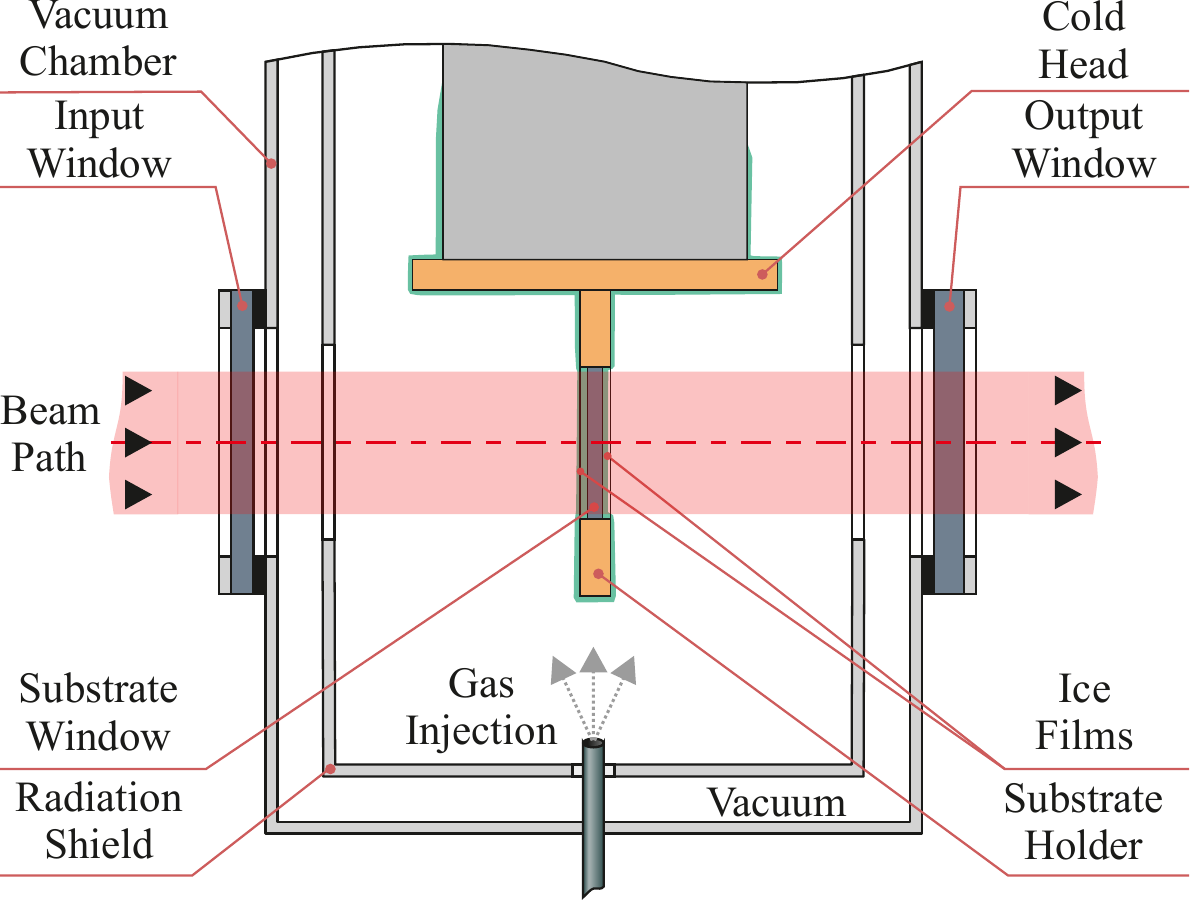}
    \caption{
    Schematic representation of the cryostat's vacuum chamber, coupled to the Bruker Vertex 70v spectrometer at CAS. Adapted from \citet{Giuliano2019}.
    }
    \label{fig:cryostat}
\end{figure}

A schematic diagram of the cryostat setup is shown in Fig.~\ref{fig:cryostat}. The cryostat is equipped with two optical ports and a gas injection port. The gas inlet, positioned approximately \SI{7}{cm} from the substrate, was not directed towards the substrate in order to ensure uniform background ice deposition on both sides. Temperature control and stabilisation were achieved using a silicon diode sensor connected to a Lake Shore Model 335 controller. High-resistivity float-zone silicon (HRFZ-Si) windows, purchased from Tydex, were used as optical interfaces. The substrate onto which the ices were deposited was made of the same material. 

The vacuum system comprised a turbomolecular pump with a nitrogen pumping speed of \SI{85}{\liter\per\second}, backed by a rotary pump with a capacity of \SI{5}{\cubic\metre\per\hour}. This configuration yielded a base pressure of approximately \SI{1e-7}{mbar} upon cooling to \SI{10}{\kelvin}. Under normal operating conditions, the minimum measurable temperature at the sample holder reached \SI{5}{\kelvin}. Further details of the cryogenic setup are provided in \citet{Giuliano2019}.

\subsection{Ice preparation}

The samples used for ice preparation included \ch{H2O} (distilled), \ch{CO} (Linde gas, $>$99\% purity), \ch{CH3OD} (methanol-OD, Sigma Aldrich, $>$99\% purity), \ch{CH2DOH} (methanol-d$_1$, Cambridge Isotope Laboratories, $>$98\% purity),  \ch{CHD2OH} (methanol-d$_2$, Sigma Aldrich, $>$98\% purity), \ch{CD3OH} (methanol-d$_3$, Thermo Fisher Scientific, $>$99.5\% purity, NMR grade), and  \ch{CD3OD} (methanol-d$_4$, Sigma Aldrich, $>$99.8\% purity). All liquid samples were purified using three freeze–pump–thaw cycles to remove volatile contaminants.

The purified gases were then mixed in a stainless-steel cylinder (Swagelok) at specified ratios, with partial pressures calculated using the ideal gas law. The gas mixture was introduced into the vacuum chamber through a \SI{6}{\milli\metre} stainless-steel pipe, controlled by a metering valve. Upon expansion into the vacuum chamber, the gas condensed onto the silicon substrate, forming an ice layer. Following initial data acquisition at \SI{10}{\kelvin}, the ice samples were subsequently heated to \SI{120}{\kelvin} at a rate of \SIrange{1}{2}{\kelvin\per\minute}. Once \SI{120}{\kelvin} was reached, the samples were held at this temperature for one hour to allow the annealing process to stabilise.

For pure methanol ices and binary mixtures containing \ch{CO}, each sample was deposited in \SI{30}{\second} increments, resulting in a total deposition time of \SI{3}{\minute}. Spectra were acquired after each incremental deposition step. After \SI{2}{\minute} of deposition, the gas mixing chamber was evacuated and refilled to restore its initial gas mixture and pressure. Under these conditions, the maximum pressure in the cryostat chamber during deposition reached approximately \SI{1e-3}{mbar}, and the sample holder temperature remained within \SI{2}{\kelvin} of its initial value.

For mixtures containing \ch{H2O}, the total deposition time was extended to \SI{10}{\minute} due to the vapour pressure of \ch{H2O} being approximately five times lower than that of methanol.
 This adjustment resulted in a lower total pressure in the mixing chamber to maintain the desired mixture ratios. The deposition process was conducted in \SI{2}{\minute} steps. To ensure stable and uniform deposition conditions, the gas mixing chamber was evacuated and refilled with the initial gas composition after \SI{6}{\minute}. Spectra were collected following each deposition step. Under these conditions, the maximum pressure in the cryostat chamber reached approximately \SI{1e-3}{mbar}, while the sample holder temperature remained stable within \SI{2}{\kelvin} of its initial value.

A summary of the ice compositions and temperatures used for each experiment is presented in Table~\ref{tab:mixing_ratios}.

\begin{table}[ht]
\renewcommand{\arraystretch}{1.25}
\caption{List of astrophysical ice analogues used in the experiments.}
\label{tab:mixing_ratios}
\begingroup
\fontsize{11}{13}\selectfont
\begin{adjustbox}{max width=\columnwidth}
\begin{tabular}{>{\raggedright\arraybackslash}p{3.0cm}
                >{\raggedright\arraybackslash}p{6.6cm}
                >{\centering\arraybackslash}p{1.6cm}}
\toprule
Ice & Composition / Mixing ratios & $T$ [K] \\
\midrule
Methanol
 & \ch{CH3OH}
 & 10, 120 \\
Deuterated Methanol
 & \ch{CH3OD}, \ch{CH2DOH}, \ch{CHD2OH}, \ch{CD3OH}, \ch{CD3OD}
 & 10, 120 \\
Binary Mixture
 & \ch{CH3OH} : $X$\newline
   Ratios: 10:1, 1:1, 1:2
 & 10, 120 \\
Ternary / \ch{H2O} Mixture 
 & \ch{H2O} : \ch{CH3OH} : $X$\newline
   Ratios: 10:9:1, 2:1:1, 3:1:2
 & 10, 120 \\
Ternary / \ch{CO} Mixture 
 & \ch{CO} : \ch{CH3OH} : $X$\newline
   Ratios: 10:9:1, 2:1:1, 3:1:2
 & 10, 120 \\
\bottomrule
\end{tabular}
\end{adjustbox}
\endgroup
\tablefoot{$X$ $ \in \{\ch{CH3OD}, \ch{CH2DOH}, \ch{CHD2OH}, \ch{CD3OH}, \ch{CD3OD}\}.$ All samples were deposited at \SI{10}{\kelvin} and subsequently annealed to \SI{120}{\kelvin}.}
\end{table}

\subsection{Film thickness}

For independent ice film thickness measurements, the cryostat unit was removed from the spectrometer’s sample compartment, and the original silicon windows were replaced with quartz windows. A He--Ne laser (\SI{633}{\nano\metre}) was directed onto each side of the substrate at incidence angles of $\theta_i=\SI{13}{\degree}$ and $\theta_i=\SI{9.5}{\degree}$ ($\Delta\theta_i = \SI{0.5}{\degree}$). Reflected beams from each side were measured using a Thorlabs powermeter. Individual thicknesses were determined by counting interference maxima on each side, and the total film thickness was obtained by summing the contributions from both sides. Table~\ref{tab:film_thickness} summarises the ice thicknesses and the average deposition rates derived for the laboratory ice mixtures. The total ice film thickness was determined using

\begin{equation}
\label{eq:thickness}
\Delta d = \frac{N \lambda}{2 n_{f}\sqrt{1-\sin^2(\theta_i) / n^2_{f}}}.
\end{equation}

This expression is derived from thin-film interference at oblique incidence. Here, $\Delta d$ represents the ice thickness, $N$ is the number of interference maxima, $\lambda$ is the wavelength of the laser source, $n_{f}$ is the refractive index of the ice film at \SI{633}{\nano\metre}, and $\theta_i$ is the angle of incidence.

To obtain precise thickness values for our mixtures, we used a FORTRAN routine that generates a theoretical interference curve to find an optimal value for $n_{f}$. The routine determines the refractive index of the film by varying the $n_{f}$ value until the modelled interference curve reproduces the amplitude of the experimental data\footnote{\url{https://oldwww.oact.inaf.it/thickness/}}. Further details on the FORTRAN routine can be found in \citet{Baratta1998, Scire2019}.

\begin{table}[ht]
\raggedright
\caption{Film thickness measurements for selected ice samples.}
\label{tab:film_thickness}
\footnotesize
\setlength{\tabcolsep}{6pt}
\renewcommand{\arraystretch}{1.05}
\begin{tabular*}{\columnwidth}{@{\extracolsep{\fill}}lcc}
\toprule
Species & Total thickness [\si{\um}] & Deposition rate [\si{\um\per\minute}] \\
\midrule
\ch{CH3OH}        & 6.79 $\pm$ 0.35 & 2.26 $\pm$ 0.12 \\
\ch{CH2DOH}       & 6.75 $\pm$ 0.35 & 2.25 $\pm$ 0.12 \\
\ch{CHD2OH}       & 6.72 $\pm$ 0.35 & 2.24 $\pm$ 0.12 \\
Binary mixture    & 6.94 $\pm$ 0.40 & 2.31 $\pm$ 0.12 \\
\ch{H2O} mixture  & 5.49 $\pm$ 0.30 & 0.55 $\pm$ 0.03 \\
\ch{CO} mixture   & 5.87 $\pm$ 0.36 & 1.96 $\pm$ 0.10 \\
\bottomrule
\end{tabular*}
\tablefoot{Thickness measurements were performed for a representative subset of experiments. For the binary, \ch{H2O}, and \ch{CO} mixtures, only \ch{CH2DOH} and \ch{CHD2OH} at 1:1 and 2:1:1 ratios were measured (Table~\ref{tab:mixing_ratios}).}
\end{table}

\subsection{Spectral data fitting}

Selected infrared spectra used for band assignment were processed using
\emph{OriginPro\,2024} software. Baseline correction was applied individually to each spectrum using the
\texttt{Quick Peak} routine, with the \emph{Baseline: Min\&Max} option selected.
This method applies a linear baseline defined by the minimum and maximum anchor points identified by the routine and was found to provide stable and reproducible baselines across the analysed spectra.

Peak positions were subsequently determined using the
\emph{Find Peak: 1st-Derivative Savitzky--Golay} algorithm, employing a
second-order polynomial and a 20-point smoothing window. Peak maxima were
identified from the zero-crossings of the first derivative. The routine
simultaneously returns peak height, full width at half maximum (FWHM), and
integrated band area following baseline subtraction. Values reported in
Tables~\ref{tab:mode1_OH_stretch}–\ref{tab:mode7_Torsion} represent averages of
three independent fitting runs, with peak positions reproducible within
$\pm1~\rcm$ for the spectra used in the band assignment.

Several vibrational regions contain overlapping bands, particularly where
multiple bending and deformation modes contribute within a narrow spectral
interval. As a result, direct and unambiguous assignment of individual features
is not always possible. To maintain clarity, a simplified
notation is therefore adopted and introduced in each subsection, enabling
systematic comparison of corresponding spectral regions among methanol isotopologues.

Spectral features are grouped according to their vibrational character, with
the analysis focusing on the strongest bands of relevance for astrophysical
applications. Very weak features, including faint overtones, combination
bands, and the lowest-frequency lattice modes, are not discussed in detail, as
they fall outside the scope of the present study.

To aid band identification, spectra recorded at \SI{10}{\kelvin} (amorphous
ice; see Fig.~\ref{fig:CH3OH_isotopologues_10K}) were compared with spectra
acquired after annealing to \SI{120}{\kelvin} (partially crystalline ice; see
Fig.~\ref{fig:CH3OH_isotopologues_120K}). Annealing leads to systematic band
sharpening and the separation of features that are blended in the amorphous
phase, facilitating comparison of overlapping vibrational modes across
isotopologues.

Under identical experimental conditions, peak positions are reproducible to
within $\pm1~\rcm$. When additional systematic effects related to baseline
treatment, ice thickness, refractive index assumptions, and uncertainties in
mixture preparation are taken into account, apparent frequency shifts of
$1$–$3~\rcm$ fall within the overall experimental uncertainty. Consequently, small spectral variations discussed in this work are interpreted comparatively rather than as absolute frequency shifts. Integrated band areas and linear correlations are used to quantify relative trends between experiments rather than to derive absolute band strengths.

\subsection{Computational details}

Quantum-chemical calculations were performed to support the assignment and interpretation of the experimental infrared spectra, with particular emphasis on overtone and combination bands that are difficult to disentangle experimentally. All calculations were carried out using Gaussian~16, Revision C.01 \citep{Gaussian2016}. Molecular geometries were optimised using the double-hybrid density functional B2PLYP \citep{Grimme2006} in combination with the correlation-consistent aug-cc-pVQZ quadruple-$\zeta$ basis set \citep{Dunning1989, Kendall1992}, employing very tight convergence criteria and an ultrafine integration grid. Harmonic and fully anharmonic vibrational analyses were subsequently performed using second-order vibrational perturbation theory (VPT2), as implemented in Gaussian \citep{Barone2005}.

This computational protocol enables explicit treatment of anharmonic effects and provides access to fundamental vibrations as well as first and second overtones and binary combination bands. The chosen level of theory is equivalent to that employed in the high-accuracy benchmark study of \citealt{Grabska2017}, and has been shown to yield reliable vibrational frequencies and intensities for methanol and related systems.

For the partially deuterated isotopologues \ch{CH2DOH} and \ch{CHD2OH}, only the lowest-energy rotamers of $C_s$ symmetry were considered. This choice reflects the dominant conformers expected under the low-temperature conditions of the experiments and avoids unnecessary duplication of near-degenerate vibrational patterns. Mode assignments were determined by combining graphical inspection of the normal-mode displacement vectors and Bayesian linear regression with automatic relevance determination using the vibrational mode automatic relevance determination (VMARD) approach \citep{Teixeira2019}. These results were further validated through a systematic comparison with established experimental and theoretical reference data, in particular the methanol vibrational atlas of \citet{Moruzzi2018} and the recent analysis of \citet{Dinu2024}.

To improve the quantitative comparison with experiment, calculated frequencies were scaled by reference to literature values compiled by \citet{Hanninen2003}. The unscaled anharmonic frequencies already show good agreement with experiment, yielding a cumulative relative deviation of 0.0986 when evaluated over the 12 fundamental modes as

\begin{equation}
\mathrm{dev} = \sum_i \left| \frac{\nu_i^{\mathrm{calc}} - \nu_i^{\mathrm{exp}}}{\nu_i^{\mathrm{exp}}} \right|.
\end{equation}

Here, $\nu_i^{\mathrm{exp}}$ and $\nu_i^{\mathrm{calc}}$ denote the experimental and calculated frequencies, respectively. Application of a uniform scaling factor of 1.003667 reduces this cumulative deviation slightly to 0.0968. In the following analysis, the scaled anharmonic frequencies are therefore adopted. Given the small magnitude of this correction, both scaled and unscaled results are considered reliable, and the calculations are used primarily to establish the relative ordering, vibrational character, and mode coupling rather than to reproduce absolute band positions.

The anharmonic calculations presented here are employed selectively to support the vibrational assignments discussed in this work. A comprehensive theoretical analysis of methanol and its isotopologues, including a systematic exploration of anharmonic effects and mode couplings, is currently in preparation and will be reported elsewhere.
 
\section{Results}
\label{sec:results}

\subsection{Pure methanol ices}
\label{sec:pure_methanol_ices}

\begin{figure*}[htbp] \centering \includegraphics[width=0.95\textwidth]{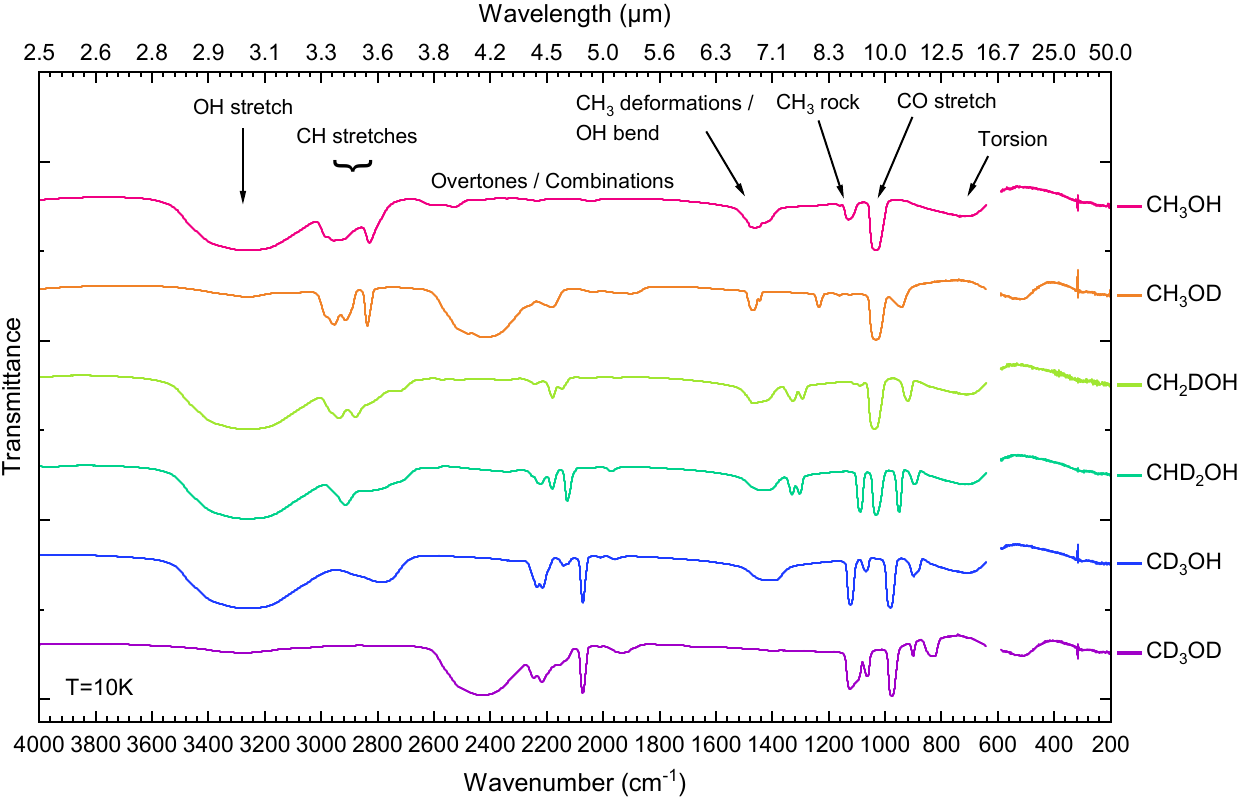}
    \caption{Transmission spectra of solid methanol and its five isotopologues at 10~K. The region between \SIrange{650}{590}{\rcm} (\SIrange{15.4}{16.9}{\um}), affected by increased noise from the beamsplitter, has been removed. The main \ch{CH3OH} vibrational modes are indicated at the top. Spectra are vertically offset for clarity.}
\label{fig:CH3OH_isotopologues_10K} 
\end{figure*} 

In this section, we analyse the vibrational spectra of pure methanol (\ch{CH3OH}) and its isotopologues by combining targeted anharmonic calculations with existing experimental and theoretical studies to assess the effects of isotopic substitution. The observed spectral shifts and changes in mode coupling arise from the altered mass distribution and symmetry properties introduced by deuteration, which significantly modify both fundamental frequencies and the structure of overtone and combination manifolds. Our assignments are therefore guided by a unified framework in which newly computed anharmonic frequencies and intensities are evaluated alongside high-resolution infrared measurements and established \textit{ab initio} results available in the literature \citep{Falk1961, Karpfen2011, Mukhopadhyay2016, Mukhopadhyay2016b, Nagaoka2007, Scire2019, Serrallach1974, Shimoaka2010}. Tentative assignments for individual vibrational modes of each deuterated isotopologue are provided in the tables presented in the main text, while only the most relevant mid-infrared bands in the 2300–800~\rcm\ region are summarised in the Appendix (Table~\ref{tab:midIR_bands}) as a focused reference for observational applications. Fig.~\ref{fig:CH3OH_isotopologues_10K} presents broadband transmission spectra of all studied isotopologues.

\subsubsection{\ch{O-H} stretch}

\begin{figure}[htbp]
    \centering
    \includegraphics[scale=0.38]{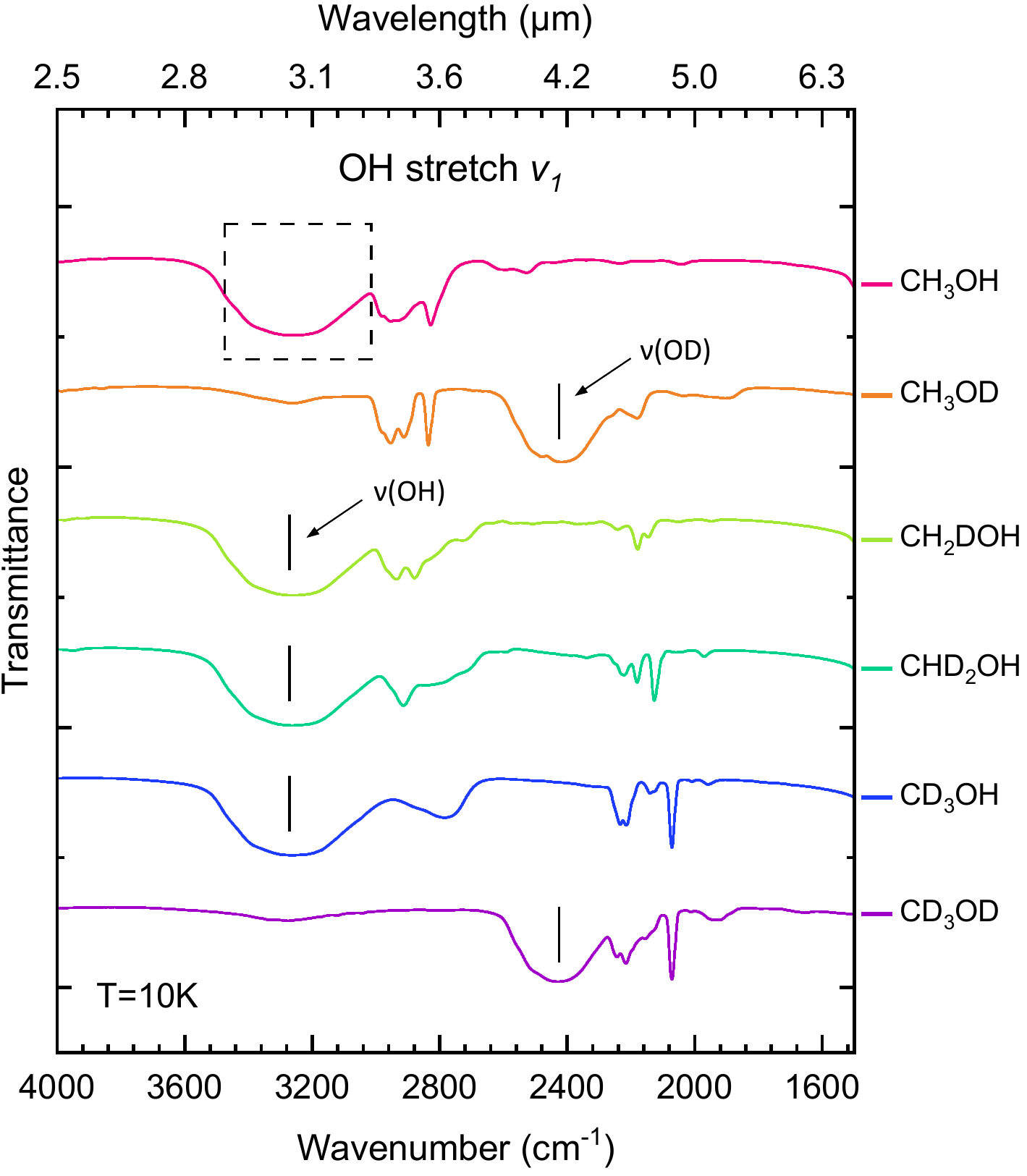}
    \caption{Transmission spectra of solid methanol and its isotopologues recorded at \SI{10}{\kelvin}, highlighting the O--H/O--D stretching region. For \ch{CH3OH}, the dashed rectangle marks the O--H stretching band. The labels $\nu(\text{O--H})$ and $\nu(\text{O--D})$ denote the fundamental O--H and O--D stretching modes, respectively. Corresponding peak positions are listed in Table~\ref{tab:mode1_OH_stretch}. Spectra are vertically offset for clarity.}
    \label{fig:mode1_OH_stretch}
\end{figure}

In amorphous \ch{CH3OH} ice, the O--H stretching band is observed at \SI{3261}{\rcm} (\SI{3.07}{\um}). Isotopologues with deuteration on the methyl group (\ch{CH2DOH}, \ch{CHD2OH}, and \ch{CD3OH}) exhibit nearly identical O--H stretching bands at 3260--3263~\rcm\ ($\approx$~\SI{3.07}{\um}), indicating that substitution in the methyl group does not significantly affect the O--H bond. In contrast, when the hydroxyl hydrogen is replaced---as in \ch{CH3OD} and \ch{CD3OD}---the O--D stretching band is shifted to lower frequencies, appearing at \SI{2418}{\rcm} (\SI{4.14}{\um}) and \SI{2430}{\rcm} (\SI{4.12}{\um}), respectively. This red--shift is consistent with the increased reduced mass of the O--D bond relative to the O--H bond. Computed fundamental $\nu$(O--H) and $\nu$(O--D) frequencies follow the same isotopic trends observed experimentally and support the assignment of the broad ice bands. Fig.~\ref{fig:mode1_OH_stretch} compares the infrared transmission spectra of pure methanol ice and its isotopologues in the O--H and O--D stretching region; Table~\ref{tab:mode1_OH_stretch} lists the band positions for the corresponding stretching modes.

\begin{table}[ht]
\caption{Fundamental hydroxyl stretches of solid methanol and its isotopologues at \SI{10}{\kelvin}.}
\label{tab:mode1_OH_stretch}
\setlength{\tabcolsep}{3pt}
\renewcommand{\arraystretch}{1.03}
\begin{adjustbox}{max width=\columnwidth}
\begin{tabular}{lcccccc}
\toprule
Mode / \si{\rcm}\,(\si{\um}) & \ch{CH3OH} & \ch{CH3OD} & \ch{CH2DOH} & \ch{CHD2OH} & \ch{CD3OH} & \ch{CD3OD} \\
\midrule
$\nu(\text{O--H})$ & 3261~(3.07) & --- & 3260~(3.07) & 3261~(3.07) & 3263~(3.06) & --- \\
$\nu(\text{O--D})$ & --- & 2418~(4.14) & --- & --- & --- & 2430~(4.12) \\
\bottomrule
\end{tabular}
\end{adjustbox}
\tablefoot{Band positions are given in \rcm, with corresponding wavelengths in \si{\um} shown in parentheses. $\nu(\text{O--H})$ and $\nu(\text{O--D})$ denote the O--H and O--D stretching vibrations, respectively. ``---'' indicates that the mode is not observed or not applicable for that isotopologue.}
\end{table}

\subsubsection{\ch{C-H} stretches}

In pure \ch{CH3OH} ice, the C--H stretching region is dominated by two prominent bands: an asymmetric stretch at \SI{2953}{\rcm} (\SI{3.39}{\um}) and a symmetric stretch at \SI{2828}{\rcm} (\SI{3.54}{\um}). These bands correspond to the fundamental C--H stretching modes of the methyl group and are in good agreement with both previous laboratory studies and the anharmonic calculations, which predict closely spaced symmetric and asymmetric C--H stretching fundamentals for solid methanol.

For \ch{CH3OD}, the asymmetric C--H stretching appears split into two distinct peaks at \SI{2954}{\rcm} and \SI{2912}{\rcm} (\SI{3.39}{\um} and \SI{3.43}{\um}), while the symmetric C--H stretch is observed at \SI{2835}{\rcm} (\SI{3.53}{\um}). Anharmonic calculations show that the fundamental C--H stretching frequencies are only marginally affected by deuteration at the hydroxyl position, suggesting that no additional C--H fundamental is expected. Instead, the extra feature observed experimentally is therefore most plausibly attributed to solid-state effects, such as site splitting and weak anharmonic mixing that lift near-degeneracies among closely spaced C--H stretching modes in the ice. Overall, deuteration at the hydroxyl position results in only minor perturbations of the C--H stretching manifold. 

In \ch{CH2DOH}, where deuteration occurs on the methyl group, the asymmetric C--H stretch shifts to \SI{2936}{\rcm} (\SI{3.41}{\um}) and the symmetric C--H stretch to \SI{2879}{\rcm} (\SI{3.47}{\um}). In addition, distinct C--D stretching bands emerge at \SI{2242}{\rcm} (\SI{4.46}{\um}), \SI{2179}{\rcm} (\SI{4.59}{\um}), and \SI{2146}{\rcm} (\SI{4.66}{\um}), providing a clear signature of deuteration. The origin of this triplet in the C–D stretching region has been discussed in terms of distinct $\nu$(C--D) contributions from different conformers and local environments in the solid phase (including possible complexation), as discussed in detail by \citet{Shimoaka2010} and \citet{Scire2019}.

\begin{figure}[htbp]
    \centering
    \includegraphics[scale=0.38]{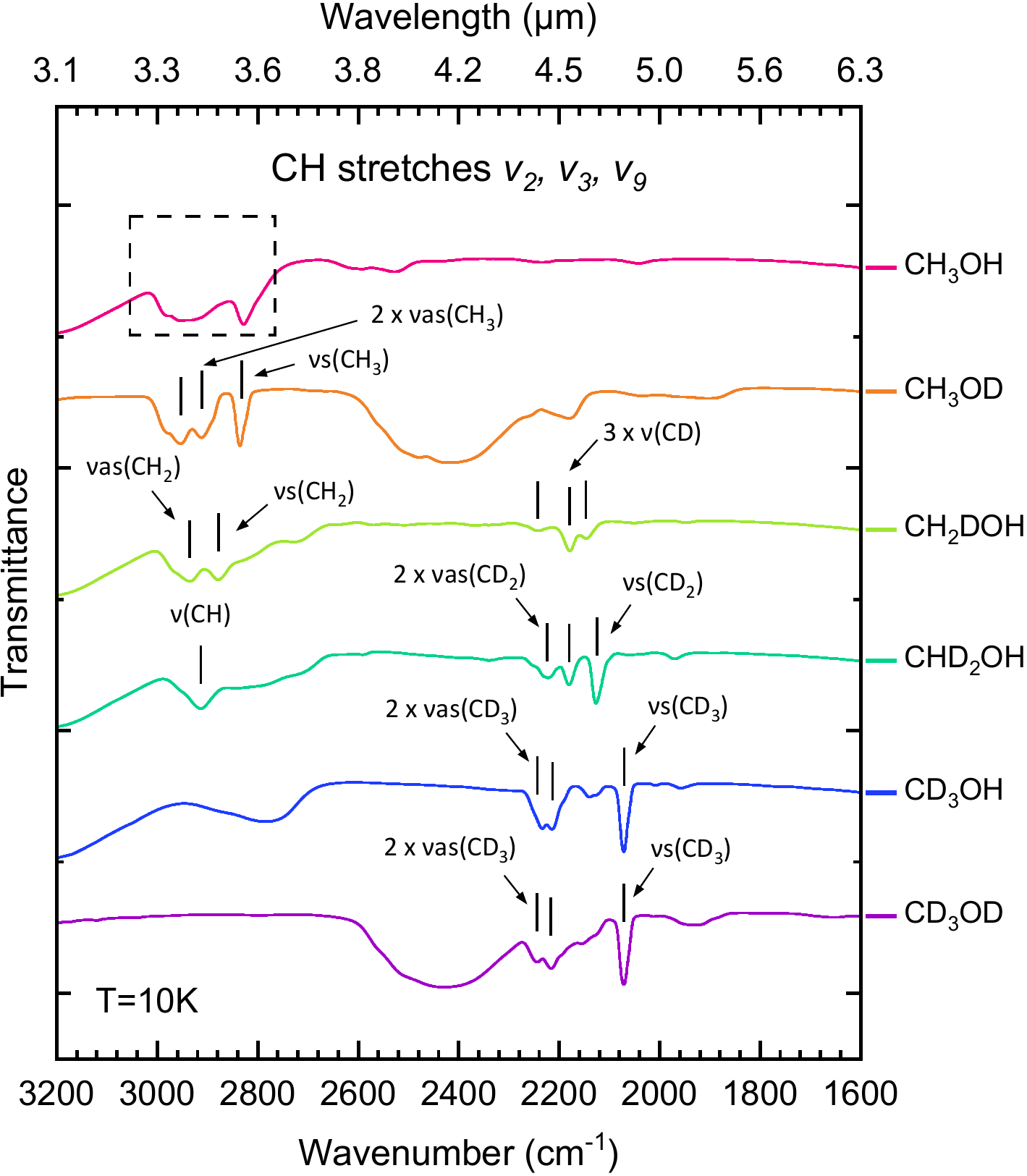}
    \caption{Transmission spectra of solid methanol and its isotopologues recorded at \SI{10}{\kelvin}, focusing on the C--H/C--D stretching region. For \ch{CH3OH}, the dashed rectangle marks the C--H stretching bands. The stretching fundamentals are labelled $\nu$as and $\nu$s, with the vibrating group indicated in parentheses. The corresponding peak positions are listed in Table~\ref{tab:mode2_CH_stretches}. Spectra are vertically offset for clarity.}
    \label{fig:mode2_CH_stretches}
\end{figure}

\begin{table}[ht]
\caption{Fundamental C--H and C--D stretches of solid methanol and its isotopologues at \SI{10}{\kelvin}.}
\label{tab:mode2_CH_stretches}
\setlength{\tabcolsep}{3pt}
\renewcommand{\arraystretch}{1}
\begin{adjustbox}{max width=\columnwidth}
\begin{tabular}{lcccccc}
\toprule
Mode / \si{\rcm}\,(\si{\um}) & \ch{CH3OH} & \ch{CH3OD} & \ch{CH2DOH} & \ch{CHD2OH} & \ch{CD3OH} & \ch{CD3OD} \\
\midrule
$\nu$as(\ch{CH3}) & 2953~(3.39) & 2954~(3.39) & --- & --- & --- & --- \\
$\nu$as(\ch{CH3}) & --- & 2912~(3.43) & --- & --- & --- & --- \\
$\nu$s(\ch{CH3})  & 2828~(3.54) & 2835~(3.53) & --- & --- & --- & --- \\
$\nu$as(\ch{CH2}) & --- & --- & 2936~(3.41) & --- & --- & --- \\
$\nu$s(\ch{CH2})  & --- & --- & 2879~(3.47) & --- & --- & --- \\
$\nu$(\ch{C-H})   & --- & --- & --- & 2913~(3.43) & --- & --- \\
$\nu$(\ch{C-D})   & --- & --- & 2242~(4.46) & --- & --- & --- \\
$\nu$(\ch{C-D})   & --- & --- & 2179~(4.59) & --- & --- & --- \\
$\nu$(\ch{C-D})   & --- & --- & 2146~(4.66) & --- & --- & --- \\
$\nu$as(\ch{CD2}) & --- & --- & --- & 2222~(4.50) & --- & --- \\
$\nu$as(\ch{CD2}) & --- & --- & --- & 2180~(4.59) & --- & --- \\
$\nu$s(\ch{CD2})  & --- & --- & --- & 2126~(4.70) & --- & --- \\
$\nu$as(\ch{CD3}) & --- & --- & --- & --- & 2233~(4.48) & 2244~(4.46) \\
$\nu$as(\ch{CD3}) & --- & --- & --- & --- & 2215~(4.51) & 2216~(4.51) \\
$\nu$s(\ch{CD3})  & --- & --- & --- & --- & 2071~(4.83) & 2071~(4.83) \\
\bottomrule
\end{tabular}
\end{adjustbox}
\tablefoot{Band positions are given in \rcm, with corresponding wavelengths in \si{\um} shown in parentheses. $\nu$ designates a stretching mode; $\nu\mathrm{as}$ and $\nu\mathrm{s}$ denote its asymmetric and symmetric components, respectively, with the vibrating group in parentheses. ``---'' indicates that the mode is not observed or not applicable for that isotopologue. Assignments are tentative.}
\end{table}

In \ch{CHD2OH}, where two of the three methyl hydrogens are replaced by deuterium, only a single C--H stretching band remains, observed at \SI{2913}{\rcm} (\SI{3.43}{\um}). The \ch{CD2} stretching region displays asymmetric bands at \SI{2222}{\rcm} (\SI{4.50}{\um}) and \SI{2180}{\rcm} (\SI{4.59}{\um}), together with a symmetric band at \SI{2126}{\rcm} (\SI{4.70}{\um}). This progressive redistribution of intensity reflects the reduced number of C--H oscillators and the increasing dominance of C--D stretching modes.

Finally, in the fully deuterated isotopologues \ch{CD3OH} and \ch{CD3OD}, C--H stretching bands are absent, and the vibrational mode shifts entirely to the C--D stretching region: for \ch{CD3OH} the asymmetric C--D stretch splits into two peaks at \SI{2233}{\rcm} (\SI{4.48}{\um}) and \SI{2215}{\rcm} (\SI{4.51}{\um}), while the symmetric C--D stretch appears at \SI{2071}{\rcm} (\SI{4.83}{\um}). \ch{CD3OD} exhibits a similar pattern, with asymmetric C--D stretching bands at \SI{2244}{\rcm} (\SI{4.46}{\um}) and \SI{2216}{\rcm} (\SI{4.51}{\um}) and a symmetric C--D stretch at \SI{2071}{\rcm} (\SI{4.83}{\um}). Fig.~\ref{fig:mode2_CH_stretches} compares the infrared transmission spectra of pure methanol ice and its isotopologues in the C--H stretching region; Table~\ref{tab:mode2_CH_stretches} lists the corresponding band positions.

\subsubsection{\ch{CH3} deformations}

In pure \ch{CH3OH}, the in-plane \ch{CH3} deformation is observed as a single broad band composed of several overlapping components, centred at \SI{1461}{\rcm} (\SI{6.84}{\um}). When the hydroxyl group is deuterated, as in \ch{CH3OD}, the \ch{CH3} deformation splits into two components, appearing at \SI{1470}{\rcm} (\SI{6.80}{\um}) and \SI{1445}{\rcm} (\SI{6.92}{\um}). 

\begin{figure}[htbp]
    \centering
    \includegraphics[scale=0.38]{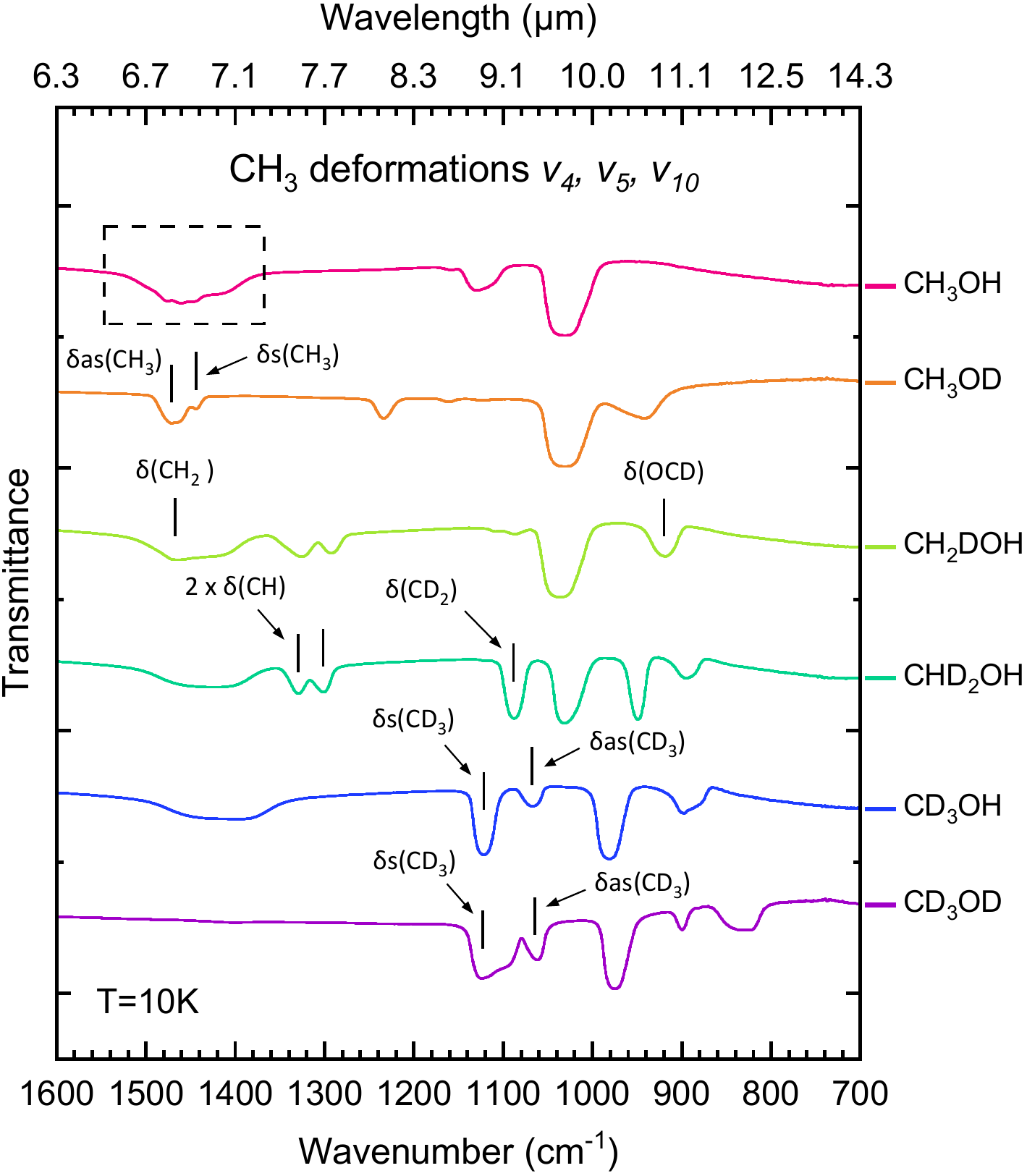}
    \caption{Transmission spectra of solid methanol and its isotopologues recorded at \SI{10}{\kelvin}, highlighting the \ch{CH3} deformation region. Modes are labelled $\delta$ for the in-plane bend. $\delta$\textnormal{as} and $\delta$\textnormal{s} denote the asymmetric and symmetric components, respectively, with the vibrating group given in parentheses. For \ch{CH3OH}, the dashed rectangle marks the \ch{CH3} deformation region. Corresponding peak positions are listed in Table~\ref{tab:mode3_CH3_deformations}. Spectra are vertically offset for clarity.}
    \label{fig:mode3_CH3_deformations}
\end{figure}

For isotopologues where deuteration occurs on the methyl group, new features emerge. \ch{CH2DOH} exhibits an in-plane \ch{CH2} deformation at \SI{1464}{\rcm} (\SI{6.83}{\um}) and a \ch{OCD} bending at \SI{919}{\rcm} (\SI{10.88}{\um}). \ch{CHD2OH} shows two C--H bending modes at \SI{1329}{\rcm} (\SI{7.52}{\um}) and \SI{1301}{\rcm} (\SI{7.69}{\um}), accompanied by a \ch{CD2} deformation band at \SI{1088}{\rcm} (\SI{9.19}{\um}). For fully deuterated species (\ch{CD3OH} and \ch{CD3OD}), the corresponding methyl deformations shift to lower frequencies, with the symmetric component appearing at approximately \SIrange{1122}{1123}{\rcm} ($\approx$ \SI{8.91}{\um}) and the asymmetric component near \SIrange{1062}{1067}{\rcm} ($\approx$ \SI{9.37}{\um}). Fig.~\ref{fig:mode3_CH3_deformations} compares the transmission spectra of pure methanol ice and its isotopologues in the \ch{CH3} deformation region; Table~\ref{tab:mode3_CH3_deformations} lists the corresponding band positions.

\begin{table}[ht]
\caption{Methyl-group deformations of solid methanol and its isotopologues at \SI{10}{\kelvin}.}
\label{tab:mode3_CH3_deformations}
\setlength{\tabcolsep}{4pt}
\renewcommand{\arraystretch}{1.05}
\begin{adjustbox}{max width=\columnwidth}
\begin{tabular}{lcccccc}
\toprule
Mode / \si{\rcm}\,(\si{\um}) & \ch{CH3OH} & \ch{CH3OD} & \ch{CH2DOH} & \ch{CHD2OH} & \ch{CD3OH} & \ch{CD3OD} \\
\midrule
$\delta$(\ch{CH3})            & 1461~(6.84) & --- & --- & --- & --- & --- \\
$\delta\mathrm{as}$(\ch{CH3}) & --- & 1470~(6.80) & --- & --- & --- & --- \\
$\delta\mathrm{s}$(\ch{CH3})  & --- & 1445~(6.92) & --- & --- & --- & --- \\
$\delta$(\ch{CH2})            & --- & --- & 1464~(6.83) & --- & --- & --- \\
$\delta$(\ch{C-H})            & --- & --- & --- & 1329~(7.52) & --- & --- \\
$\delta$(\ch{C-H})            & --- & --- & --- & 1301~(7.69) & --- & --- \\
$\delta$(\ch{OCD})            & --- & --- & 919~(10.88) & --- & --- & --- \\
$\delta$(\ch{CD2})            & --- & --- & --- & 1088~(9.19) & --- & --- \\
$\delta\mathrm{s}$(\ch{CD3})  & --- & --- & --- & --- & 1122~(8.91) & 1123~(8.90) \\
$\delta\mathrm{as}$(\ch{CD3}) & --- & --- & --- & --- & 1067~(9.37) & 1062~(9.42) \\
\bottomrule
\end{tabular}
\end{adjustbox}
\tablefoot{Band positions are given in \rcm, with corresponding wavelengths in \si{\um} shown in parentheses. $\delta$ denotes an in-plane bending mode; $\delta\mathrm{as}$ and $\delta\mathrm{s}$ designate the asymmetric and symmetric components, respectively, with the vibrating group given in parentheses. ``---'' indicates that the mode is not observed or not applicable for that isotopologue. Assignments are tentative.}
\end{table}

\subsubsection{\ch{O-H} bending}

\begin{figure}[htbp]
    \centering
    \includegraphics[scale=0.38]{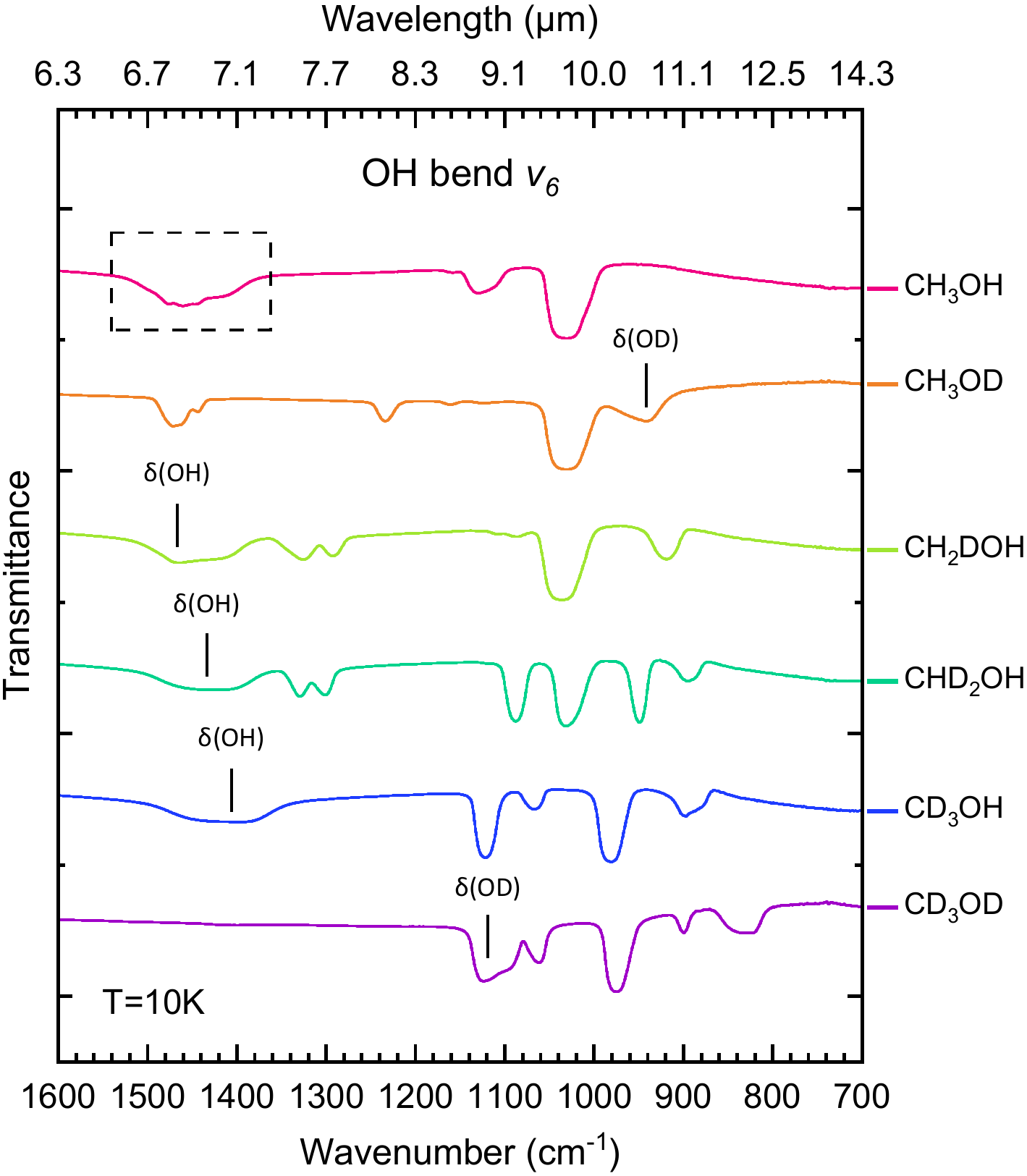}
    \caption{Transmission spectra of solid methanol and its isotopologues recorded at \SI{10}{\kelvin}, showing the O--H/O--D bending region. The bending fundamentals are labelled $\delta(\text{OH})$ and $\delta(\text{OD})$. For \ch{CH3OH}, the dashed rectangle marks the O--H bending band. Corresponding peak positions are listed in Table~\ref{tab:mode4_OH_bending}. Spectra are vertically offset for clarity.}
    \label{fig:mode4_OH_bending}
\end{figure}

The O--H bending mode in \ch{CH3OH} ice overlaps with the \ch{CH3} deformation features and appears as a broad and relatively weak band owing to strong hydrogen bonding among neighbouring molecules. For isotopologues with deuteration on the methyl group, the O--H bending mode remains blended with deformation features and appears near \SI{1464}{\rcm} (\SI{6.83}{\um}) in \ch{CH2DOH}, at \SI{1424}{\rcm} (\SI{7.02}{\um}) in \ch{CHD2OH}, and at \SI{1398}{\rcm} (\SI{7.15}{\um}) in \ch{CD3OH}. In contrast, for isotopologues with substitution at the hydroxyl position—namely, \ch{CH3OD} and \ch{CD3OD}—the O--H bending mode is replaced by an O--D bending mode, observed at \SI{942}{\rcm} (\SI{10.62}{\um}) in \ch{CH3OD} and at \SI{1123}{\rcm} (\SI{8.90}{\um}) in \ch{CD3OD}. Fig.~\ref{fig:mode4_OH_bending} compares the infrared transmission spectra of pure methanol ice and its isotopologues in the O--H bending region; Table~\ref{tab:mode4_OH_bending} lists the corresponding band positions.

\begin{table}[ht]
\caption{Hydroxyl bending fundamentals of solid methanol and its isotopologues at \SI{10}{\kelvin}.}
\label{tab:mode4_OH_bending}
\setlength{\tabcolsep}{3pt}
\renewcommand{\arraystretch}{1.03}
\begin{adjustbox}{max width=\columnwidth}
\begin{tabular}{lcccccc}
\toprule
Mode / \si{\rcm}\,(\si{\um}) & \ch{CH3OH} & \ch{CH3OD} & \ch{CH2DOH} & \ch{CHD2OH} & \ch{CD3OH} & \ch{CD3OD} \\
\midrule
$\delta(\text{O--H})$ & 1461~(6.84) & --- & 1464~(6.83) & 1424~(7.02) & 1398~(7.15) & --- \\
$\delta(\text{O--D})$ & --- & 942~(10.62) & --- & --- & --- & 1123~(8.90) \\
\bottomrule
\end{tabular}
\end{adjustbox}
\tablefoot{Band positions are given in \rcm, with corresponding wavelengths in \si{\um} shown in parentheses. $\delta(\text{O--H})$ and $\delta(\text{O--D})$ denote the O--H and O--D bending vibrations, respectively. ``---'' indicates that the mode is not observed or not applicable for that isotopologue. Assignments are tentative.}
\end{table}

\subsubsection{\ch{CH3} rocking}

In pure \ch{CH3OH} ice, the \ch{CH3} rocking modes appear as two bands at \SI{1158}{\rcm} (\SI{8.64}{\um}) and \SI{1129}{\rcm} (\SI{8.86}{\um}). For \ch{CH3OD}, these modes are observed at \SI{1161}{\rcm} (\SI{8.61}{\um}) and \SI{1123}{\rcm} (\SI{8.90}{\um}), and an additional feature at \SI{1234}{\rcm} (\SI{8.10}{\um})  also appears (previously reported by \citealt{Falk1961}), suggesting a subtle shift due to coupling with the O--H bending vibration.

In \ch{CH2DOH}, where deuteration occurs on the methyl group, the \ch{CH2} out-of-plane modes appear as a wagging band at \SI{1326}{\rcm} (\SI{7.54}{\um}), and a twisting band at \SI{1293}{\rcm} (\SI{7.73}{\um}). Two in-plane rocking components are observed at \SI{1108}{\rcm} (\SI{9.03}{\um}) and \SI{1087}{\rcm} (\SI{9.20}{\um}). In \ch{CHD2OH}, these \ch{CH3}-specific modes are no longer clearly visible, likely due to overlap with \ch{CD2} bending features.

\begin{figure}[htbp]
    \centering
    \includegraphics[scale=0.38]{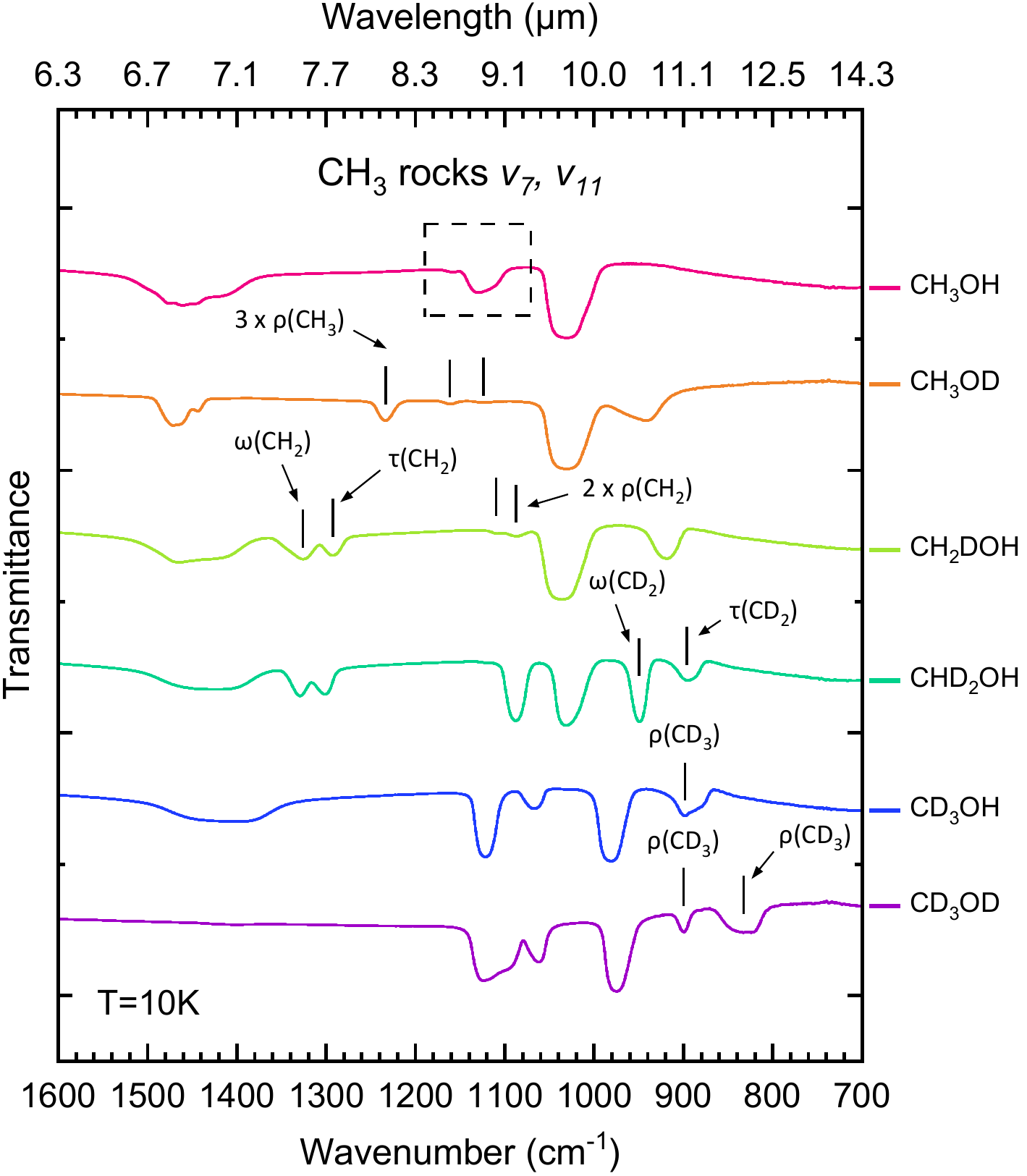}
    \caption{Transmission spectra of solid methanol and its isotopologues recorded at \SI{10}{\kelvin}, highlighting the methyl rocking region. Rocking ($\rho$), twisting ($\tau$), and wagging ($\omega$) modes are indicated, with the vibrating group given in parentheses. For \ch{CH3OH}, the dashed rectangle marks the \ch{CH3} rocking bands. Corresponding peak positions are listed in Table~\ref{tab:mode5_CH3_rocking}. Spectra are vertically offset for clarity.}
    \label{fig:mode5_CH3_rocking}
\end{figure}

Finally, among the highly deuterated isotopologues, the triply deuterated \ch{CD3OH} exhibits a \ch{CD3} rocking mode at \SI{898}{\rcm} (\SI{11.14}{\um}), whereas the fully deuterated \ch{CD3OD} shows two rocking bands at \SI{900}{\rcm} (\SI{11.11}{\um}) and \SI{832}{\rcm} (\SI{12.02}{\um}), consistent with \citet{Nagaoka2007}. Fig.~\ref{fig:mode5_CH3_rocking} compares the transmission spectra of pure methanol ice and its isotopologues in the \ch{CH3} rocking region; Table~\ref{tab:mode5_CH3_rocking} lists the corresponding band positions.

\begin{table}[ht]
\caption{Methyl rocking fundamentals of solid methanol and its isotopologues at \SI{10}{\kelvin}.}
\label{tab:mode5_CH3_rocking}
\setlength{\tabcolsep}{3pt}
\renewcommand{\arraystretch}{1.03}
\begin{adjustbox}{max width=\columnwidth}
\begin{tabular}{lcccccc}
\toprule
Mode / \si{\rcm}\,(\si{\um}) & \ch{CH3OH} & \ch{CH3OD} & \ch{CH2DOH} & \ch{CHD2OH} & \ch{CD3OH} & \ch{CD3OD} \\
\midrule
$\rho$(\ch{CH3})   & --- & 1234~(8.10) & --- & --- & --- & --- \\
$\rho$(\ch{CH3})   & 1158~(8.64) & 1161~(8.61) & --- & --- & --- & --- \\
$\rho$(\ch{CH3})   & 1129~(8.86) & 1123~(8.90) & --- & --- & --- & --- \\
$\omega$(\ch{CH2}) & --- & --- & 1326~(7.54) & --- & --- & --- \\
$\tau$(\ch{CH2})   & --- & --- & 1293~(7.73) & --- & --- & --- \\
$\rho$(\ch{CH2})   & --- & --- & 1108~(9.03) & --- & --- & --- \\
$\rho$(\ch{CH2})   & --- & --- & 1087~(9.20) & --- & --- & --- \\
$\omega$(\ch{CD2}) & --- & --- & --- & 949~(10.54) & --- & --- \\
$\tau$(\ch{CD2})   & --- & --- & --- & 895~(11.17) & --- & --- \\
$\rho$(\ch{CD3})   & --- & --- & --- & --- & 898~(11.14) & 900~(11.11) \\
$\rho$(\ch{CD3})   & --- & --- & --- & --- & --- & 832~(12.02) \\
\bottomrule
\end{tabular}
\end{adjustbox}
\tablefoot{Band positions are given in \rcm, with corresponding wavelengths in \si{\um} shown in parentheses. $\rho$ denotes in--plane rocking, $\omega$ out--of--plane wagging, and $\tau$ out--of--plane twisting; the vibrating group is indicated in parentheses. ``---'' indicates that the mode is not observed or not applicable for that isotopologue. Assignments are tentative.}
\end{table}

\subsubsection{\ch{C-O} stretch}

\begin{figure}[htbp]
    \centering
    \includegraphics[scale=0.38]{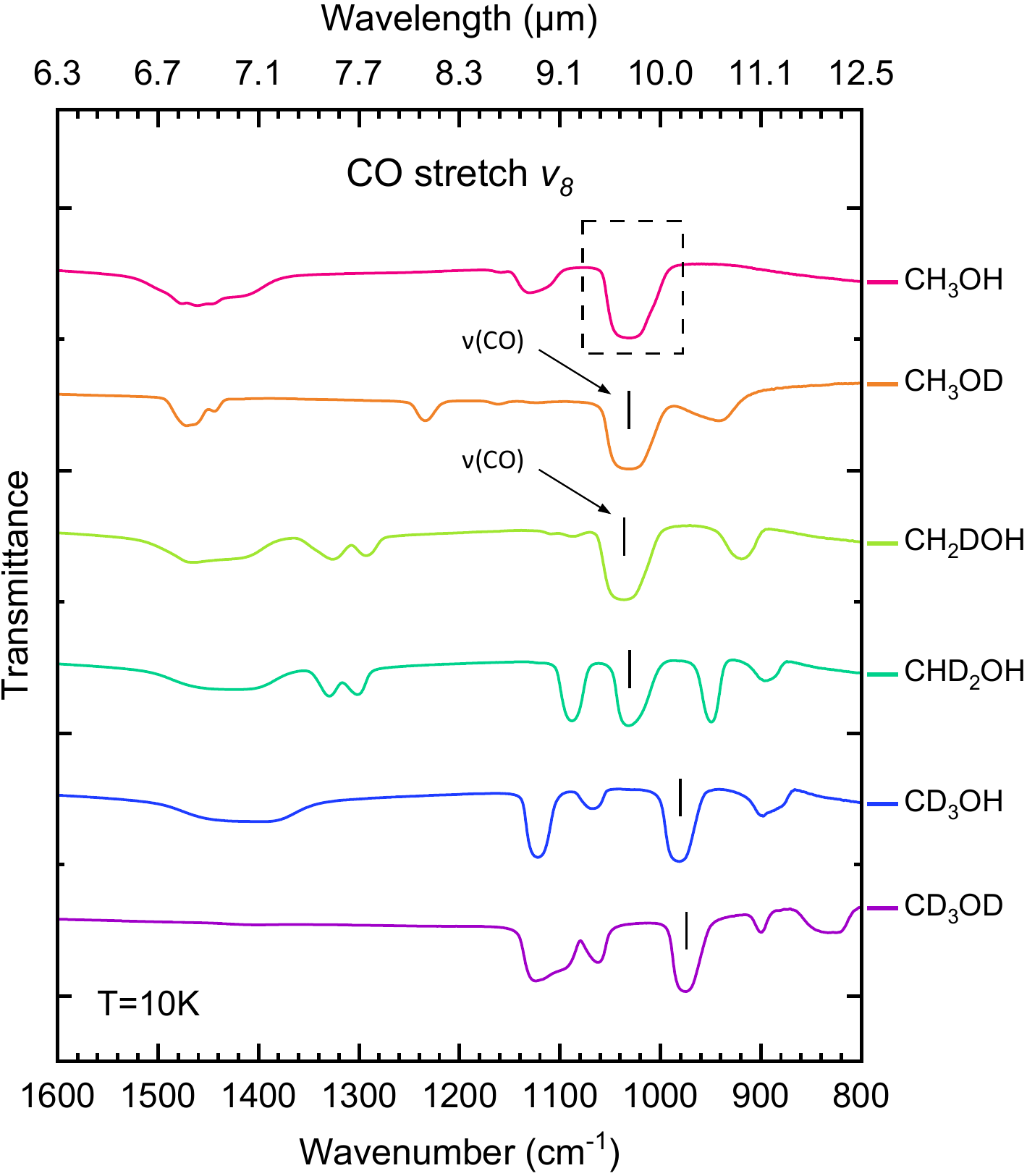}
    \caption{Transmission spectra of solid methanol and its isotopologues recorded at \SI{10}{\kelvin}, highlighting the C--O stretching region. The fundamental is labelled $\nu(\text{C--O})$. For \ch{CH3OH}, the dashed rectangle marks the C--O stretching band. Corresponding peak positions are listed in Table~\ref{tab:mode6_CO_stretch}. Spectra are vertically offset for clarity.}
    \label{fig:mode6_CO_stretch}
\end{figure}

In pure \ch{CH3OH} ice, the C--O stretching band is observed at \SI{1031}{\rcm} (\SI{9.70}{\um}). For hydroxyl-deuterated \ch{CH3OD}, the C--O stretch remains unchanged at \SI{1031}{\rcm}. Among the methyl-deuterated isotopologues, \ch{CHD2OH} also exhibits the C~--~O stretching band at \SI{1031}{\rcm}, whereas \ch{CH2DOH} shows a modest shift to slightly higher wavenumber at \SI{1036}{\rcm} (\SI{9.65}{\um}). By contrast, a clear red--shift is observed for the more heavily deuterated isotopologues, with the C--O stretch appearing at \SI{981}{\rcm} (\SI{10.19}{\um}) in \ch{CD3OH} and at \SI{975}{\rcm} (\SI{10.26}{\um}) in \ch{CD3OD}. These values are consistent with previous vapour-phase observations \citep{Falk1961} and indicate that substitution at the hydroxyl group or partial deuteration of the methyl group results in only small, non-systematic changes in the C--O stretching frequency, whereas extensive ($\geq 3$ H$\rightarrow$D) substitution leads to a systematic shift towards lower frequencies. As with the O--H stretching region, the deposition conditions were adjusted to enhance saturation of the C--O band, thereby improving the detection of weaker mid-infrared and far-infrared features. Fig.~\ref{fig:mode6_CO_stretch} compares the transmission spectra of pure methanol ice and its isotopologues in the C--O stretching region; Table~\ref{tab:mode6_CO_stretch} lists the corresponding band positions.

\begin{table}[ht]
\caption{C--O stretching fundamental of solid methanol and its isotopologues at \SI{10}{\kelvin}.}
\label{tab:mode6_CO_stretch}
\setlength{\tabcolsep}{3pt}
\renewcommand{\arraystretch}{1.03}
\begin{adjustbox}{max width=\columnwidth}
\begin{tabular}{lcccccc}
\toprule
Mode / \si{\rcm}\,(\si{\um}) & \ch{CH3OH} & \ch{CH3OD} & \ch{CH2DOH} & \ch{CHD2OH} & \ch{CD3OH} & \ch{CD3OD} \\
\midrule
$\nu(\text{C--O})$ & 1031~(9.70) & 1031~(9.70) & 1036~(9.65) & 1031~(9.70) & 981~(10.19) & 975~(10.26) \\
\bottomrule
\end{tabular}
\end{adjustbox}
\tablefoot{Band positions are given in \rcm, with corresponding wavelengths in \si{\um} shown in parentheses. $\nu(\text{C--O})$ denotes the C--O stretching vibration. Values reflect shifts due to isotopic substitution.}
\end{table}

\subsubsection{Torsion}

In pure \ch{CH3OH} ice, the torsional mode, which involves the internal rotation of the methyl group about the C--O bond, is observed at \SI{710}{\rcm} (\SI{14.1}{\um}). For isotopologues deuterated on the methyl group (\ch{CH2DOH}, \ch{CHD2OH}, and \ch{CD3OH}), the torsional frequency remains close to that of \ch{CH3OH}. In contrast, substitution at the hydroxyl position, as in \ch{CH3OD} and \ch{CD3OD}, results in a significant red--shift, with the torsional mode appearing at \SI{520}{\rcm} (\SI{19.2}{\um}) and \SI{518}{\rcm} (\SI{19.3}{\um}), respectively.

Weak lattice phonon features are already discernible in amorphous \ch{CH3OH} ice at \SI{10}{\kelvin} ($\approx$~\SI{307}{\rcm} and $\approx$~\SI{243}{\rcm}). After annealing to \SI{120}{\kelvin}, the crystalline phase exhibits a well-resolved triplet at \SI{348}{\rcm}, \SI{293}{\rcm}, and \SI{237}{\rcm}. The same three bands appear for every deuterated isotopologue, each shifting by no more than \SI{10}{\rcm} relative to \ch{CH3OH}. A more detailed comparison of these torsional and lattice modes is provided in Appendix~\ref{app:lattice_modes}. Fig.~\ref{fig:mode7_Torsion} compares the infrared transmission spectra of pure \ch{CH3OH} ice and its isotopologues in the torsional region; Table~\ref{tab:mode7_Torsion} lists the corresponding band positions.

\begin{table}[ht]
\caption{Torsional fundamental of solid methanol and its isotopologues at \SI{10}{\kelvin}.}
\label{tab:mode7_Torsion}
\setlength{\tabcolsep}{3pt}
\renewcommand{\arraystretch}{1.03}
\begin{adjustbox}{max width=\columnwidth}
\begin{tabular}{lcccccc}
\toprule
Mode / \si{\rcm}\,(\si{\um}) & \ch{CH3OH} & \ch{CH3OD} & \ch{CH2DOH} & \ch{CHD2OH} & \ch{CD3OH} & \ch{CD3OD} \\
\midrule
$\Phi(\text{C--O})$ & 710~(14.1) & 520~(19.2) & 710~(14.1) & 710~(14.1) & 707~(14.1) & 518~(19.3) \\
\bottomrule
\end{tabular}
\end{adjustbox}
\tablefoot{Band positions are given in \rcm, with corresponding wavelengths in \si{\um} shown in parentheses. $\Phi(\text{C--O})$ denotes the C--O torsional mode. Values reflect shifts due to isotopic substitution.}
\end{table}

\begin{figure}[htbp]
    \centering
    \includegraphics[scale=0.38]{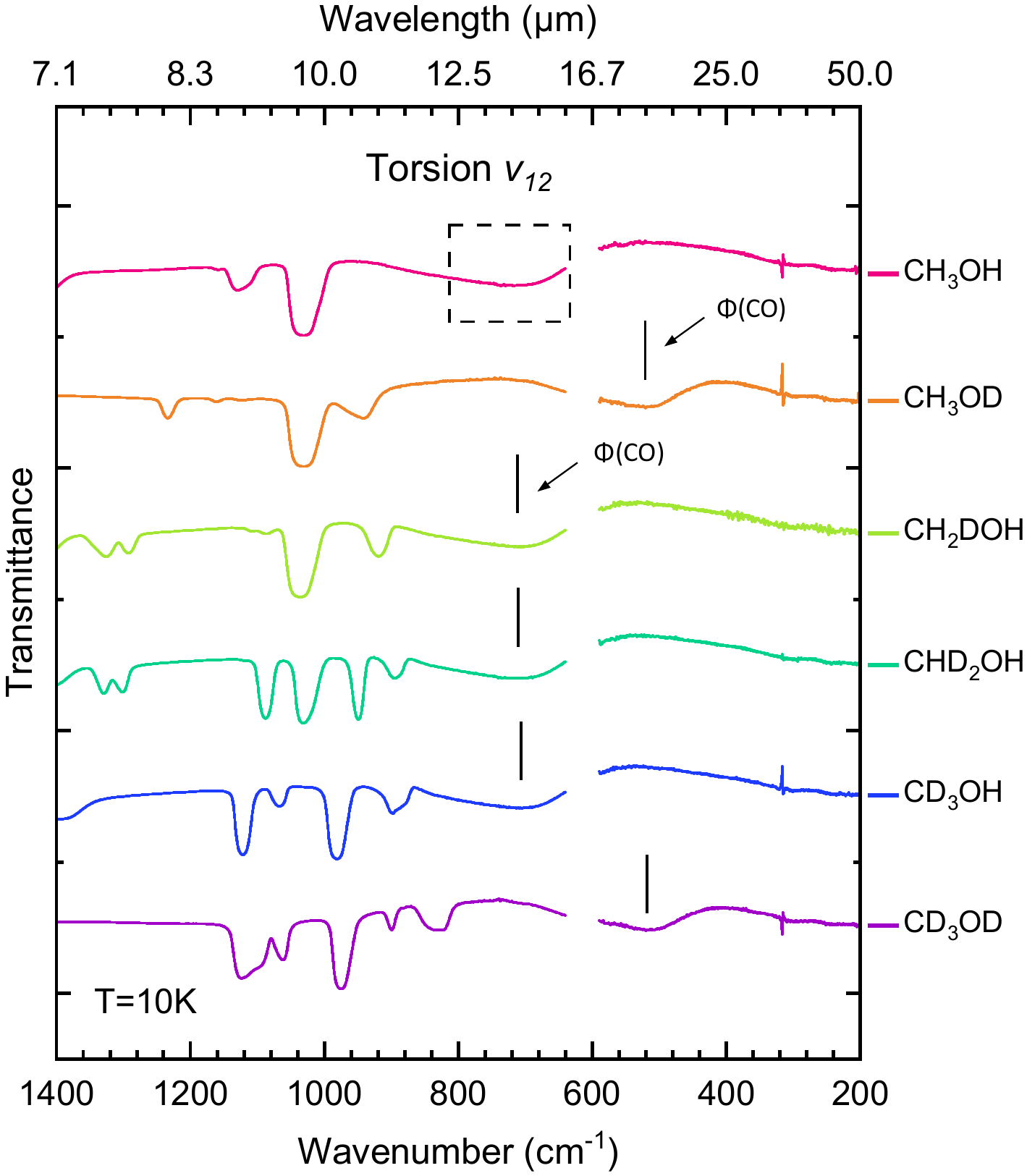}
    \caption{Transmission spectra of solid methanol and its isotopologues recorded at \SI{10}{\kelvin}, highlighting the methyl-torsion region. The torsional fundamental is labelled $\Phi(\text{C--O})$. For \ch{CH3OH}, the dashed rectangle marks the torsional band. Corresponding peak positions are listed in Table~\ref{tab:mode7_Torsion}. Spectra are vertically offset for clarity.}
    \label{fig:mode7_Torsion}
\end{figure}

\subsubsection{Overtone and combination bands}

\begin{figure}[t!]
    \centering
    \includegraphics[scale=0.38]{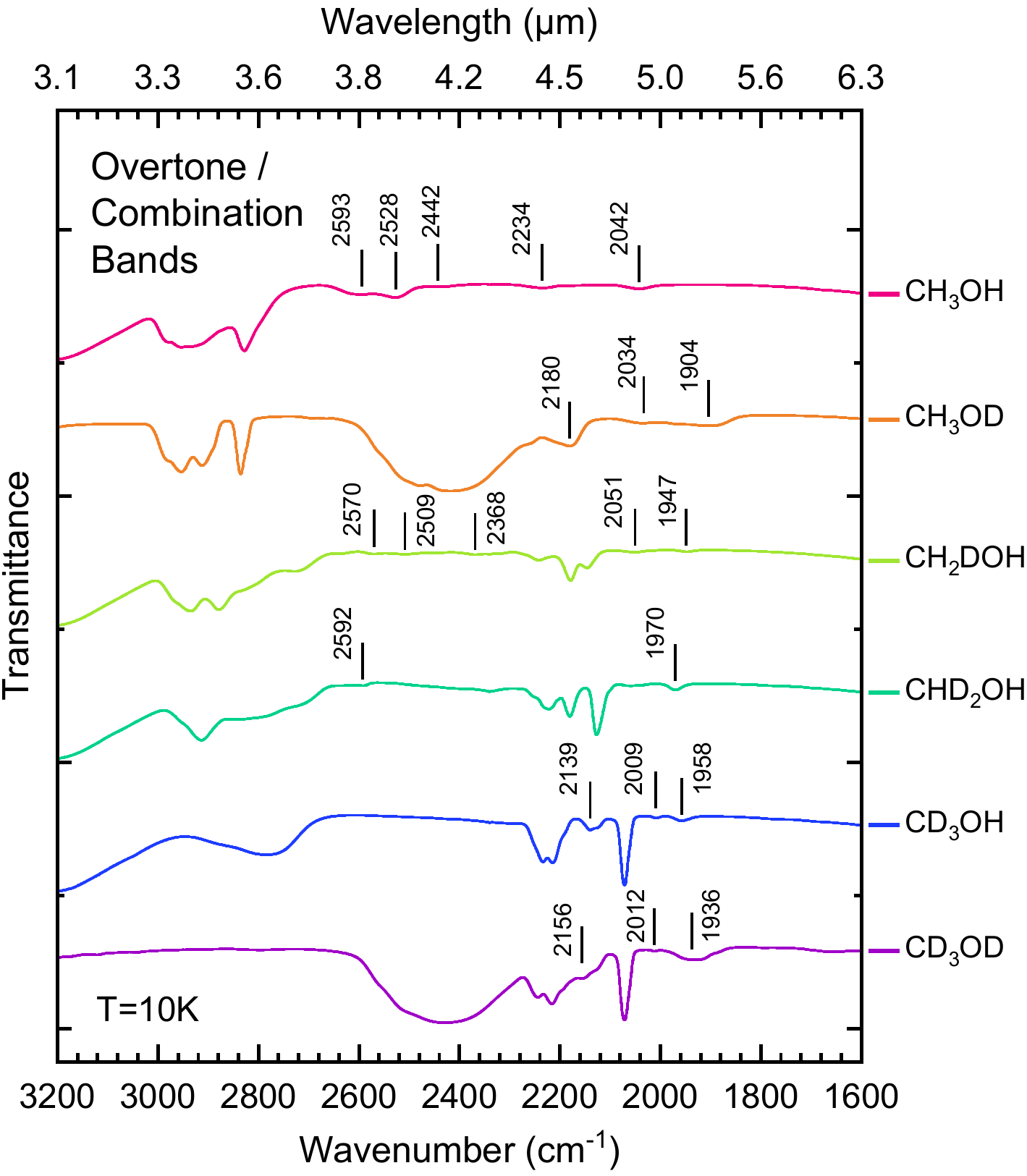}
    \caption{Transmission spectra of solid methanol and its isotopologues recorded at \SI{10}{\kelvin}, showing combination and overtone bands. Spectra are vertically offset for clarity.}
    \label{fig:mode8_Overtones}
\end{figure}

Fig.~\ref{fig:mode8_Overtones} compares the infrared transmission spectra of pure \ch{CH3OH} ice and its isotopologues in the \SIrange{2600}{1900}{\rcm} (\SIrange{3.85}{5.26}{\um}) region, which is dominated by overtone and combination transitions. For isotopologues containing C--D bonds, this window also includes fundamental C--D stretching modes. Across all isotopologues, the dominant and most robust feature in this region is the first overtone of the C--O stretching mode, $2\nu(\mathrm{CO})$, appearing between 1936 and 2051~\rcm (5.17 and 4.86~\um) depending on isotopic substitution. Anharmonic calculations identify this transition as the strongest contributor in the overtone/combination window. All other features arise from clusters of weak, strongly mixed combination bands involving C--O~stretching, hydroxyl bending, and methyl or deuterated-group deformations. These bands are intrinsically weak, broadened by solid-state effects, and vary in relative intensity among isotopologues and are therefore not expected to provide primary astrophysical diagnostics. Tabulated summaries of these bands are provided in Appendix~\ref{app:NIR_bands}, where near-infrared positions are compared with literature data from \citet{Grabska2017} and mid-infrared assignments are guided by the anharmonic calculations presented in this work. Owing to the weak intensities and strong mode mixing in this region, a complete assignment of all contributing transitions is beyond the scope of the present study.

\subsection{Astrophysical ices: Binary mixtures and \ch{H2O}- and \ch{CO}-rich matrices}
\label{sec:mixtures}

In this section, we compare the infrared transmission spectra of different astrophysical ice mixtures for each deuterated methanol isotopologue, focusing on the frequency window between \SIrange{2300}{800}{\rcm} (\SIrange{4.35}{12.50}{\um}). This range was selected for several reasons. First, it corresponds to the spectral coverage of JWST/MIRI. Second, it includes vibrational regions where deuteration introduces characteristic spectral changes—namely, the appearance of shifted C--D stretching modes between \SIrange{2300}{2000}{\rcm} (\SIrange{4.35}{5}{\um}) and shifted \ch{CH3} deformation modes between \SIrange{1500}{800}{\rcm} (\SIrange{6.66}{12.5}{\um}).

All spectra were recorded after deposition at \SI{10}{\kelvin}. Each subsection presents the spectral data in four environments: pure methanol ice, a binary mixture, an \ch{H2O} matrix, and a \ch{CO} matrix (see Table~\ref{tab:mixing_ratios}). The key vibrational features of each isotopologue are highlighted under these distinct astrophysical conditions.

At the end of each subsection, a summary table lists the highlighted band positions, facilitating comparison across different environments. Notably, the characteristic bands of the deuterated species exhibit only minor frequency shifts, within the experimental uncertainty, indicating that their fundamental vibrational modes are largely unaffected by the surrounding matrix composition.

\subsubsection{\ch{CH3OD} mixtures}

\begin{figure}[htbp]
    \centering
    \includegraphics[scale=0.4]{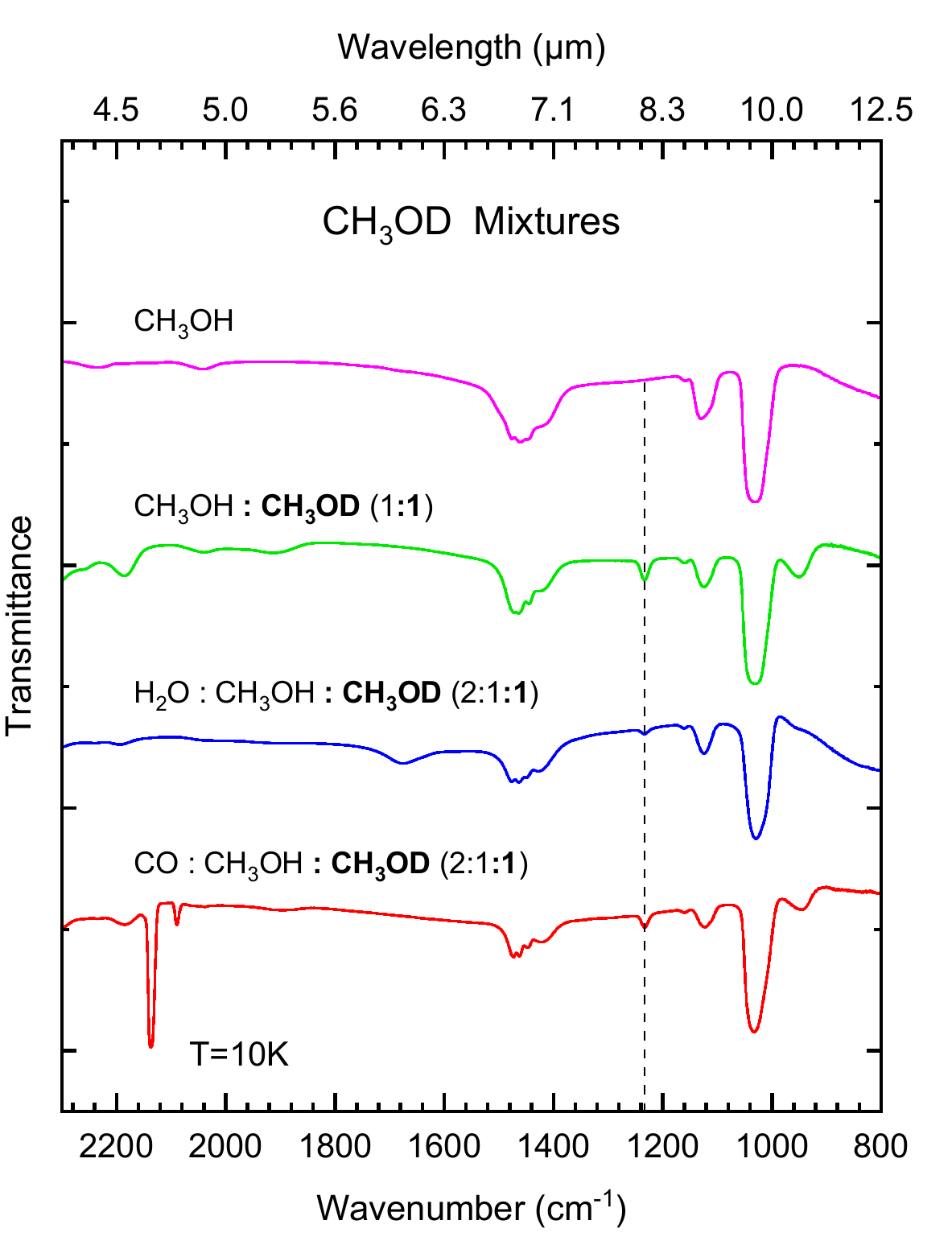}
    \caption{Transmission spectra of pure methanol ice and deuterated methanol (\ch{CH3OD}) embedded in three astrophysical ice matrices at \SI{10}{\kelvin}. Key vibrational features associated with deuteration are marked by vertical dashed lines; their centres vary within 1~\rcm\ across the matrices (see Table~\ref{tab:mixtures_CH3OD_10K}). Spectra are vertically offset for clarity.}
    \label{fig:mixtures_CH3OD_10K}
\end{figure}

For ice mixtures containing \ch{CH3OD}, a characteristic \ch{CH3}-rocking band appears at \SI{1234}{\rcm} (\SI{8.10}{\um}) in both the binary and \ch{H2O}-rich mixtures. In the \ch{CO}-rich ice, this feature shifts marginally to \SI{1233}{\rcm} (\SI{8.11}{\um}). This band is absent in pure \ch{CH3OH} ice and therefore serves as a clear spectroscopic marker of \ch{CH3OD} that is preserved across the different ice environments. The observed variation in peak position is limited to \SI{1}{\rcm} and thus lies within the experimental uncertainty. Fig.~\ref{fig:mixtures_CH3OD_10K} shows the transmission spectra of \ch{CH3OD}-containing mixtures overlaid with that of pure methanol ice, illustrating the spectral change introduced by deuteration; Table~\ref{tab:mixtures_CH3OD_10K} summarises the frequencies of the key deuterated bands highlighted in the figure for each matrix.

\begin{table}[ht]
\caption{Deuterated vibrational modes of \ch{CH3OD} in astrophysical ice mixtures at \SI{10}{\kelvin} in the mid-infrared region.}
\label{tab:mixtures_CH3OD_10K}
\setlength{\tabcolsep}{4pt}
\renewcommand{\arraystretch}{1.05}
\begin{tabular*}{\columnwidth}{@{\extracolsep{\fill}}lccc}
\toprule
Mode / \si{\rcm}\,(\si{\um}) & Binary mixture & \ch{H2O} matrix & \ch{CO} matrix \\
\midrule
\ch{CH3} rock & 1234~(8.10) & 1234~(8.10) & 1233~(8.11) \\
\bottomrule
\end{tabular*}
\tablefoot{Band positions are given in \rcm, with corresponding wavelengths in \si{\um} shown in parentheses. Minor shifts of 1~\rcm~indicate preservation of deuterated features across different ice matrices.}
\end{table}

\subsubsection{\ch{CH2DOH} mixtures}

\begin{figure}[htbp]
    \centering
    \includegraphics[scale=0.4]{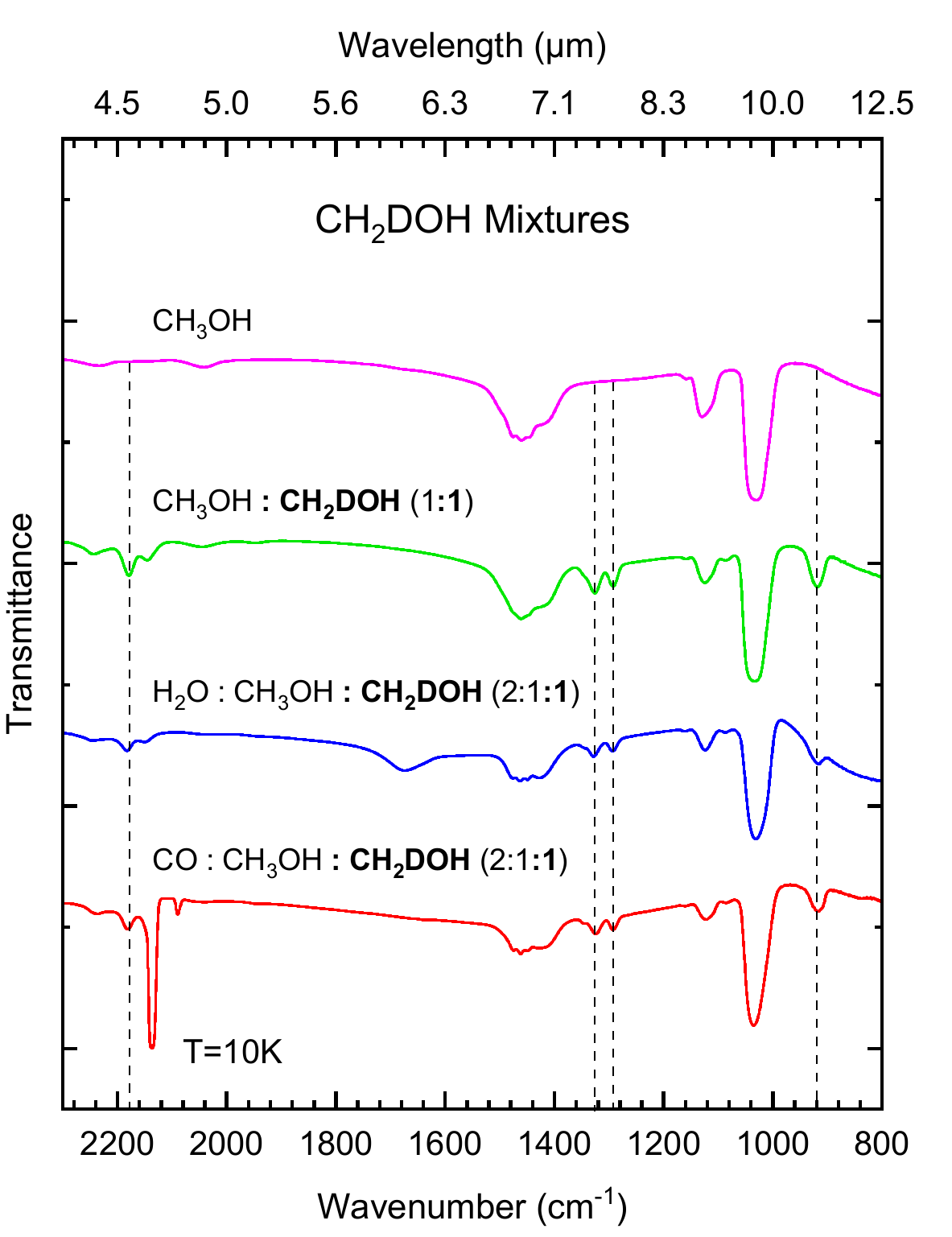}
    \caption{Transmission spectra of pure methanol ice and deuterated methanol (\ch{CH2DOH}) embedded in three astrophysical ice matrices at \SI{10}{\kelvin}. Key vibrational features associated with deuteration are marked by vertical dashed lines; their centres vary by 1--4~\rcm\ across the matrices (see Table~\ref{tab:mixtures_CH2DOH_10K}). Spectra are vertically offset for clarity.}
    \label{fig:mixtures_CH2DOH_10K}
\end{figure}

For ice mixtures containing \ch{CH2DOH}, a characteristic doublet is observed in the binary ice at \SI{1326}{\rcm} (\SI{7.54}{\um}, \ch{CH2} wag) and \SI{1293}{\rcm} (\SI{7.73}{\um}, \ch{CH2} twist). In the \ch{H2O}-rich ice, these bands shift slightly to \SI{1329}{\rcm} (\SI{7.52}{\um}) and \SI{1294}{\rcm} (\SI{7.73}{\um}), respectively, while in the \ch{CO}-rich ice they appear at \SI{1325}{\rcm} (\SI{7.55}{\um}) and \SI{1293}{\rcm} (\SI{7.73}{\um}). The C--D stretching mode is located at \SI{2179}{\rcm} (\SI{4.59}{\um}) in the binary mixture, shifting to \SI{2183}{\rcm} (\SI{4.58}{\um}) in the \ch{H2O} matrix and to \SI{2181}{\rcm} (\SI{4.59}{\um}) in the \ch{CO} matrix. A weaker \ch{OCD} bending mode is observed at \SI{919}{\rcm} (\SI{10.88}{\um}) in the binary ice, shifting to \SI{917}{\rcm} (\SI{10.90}{\um}) in the \ch{H2O}-rich mixture and to \SI{920}{\rcm} (\SI{10.87}{\um}) in the \ch{CO}-rich mixture. All observed shifts are within a few \si{\rcm} and thus lie within the overall experimental uncertainty. Fig.~\ref{fig:mixtures_CH2DOH_10K} shows the transmission spectra of \ch{CH2DOH}-containing mixtures overlaid with that of pure methanol ice, allowing comparison of the spectral changes induced by deuteration; Table~\ref{tab:mixtures_CH2DOH_10K} summarises the frequencies of the key deuterated bands highlighted in the figure for each matrix.

\begin{table}[ht]
\caption{Deuterated vibrational modes of \ch{CH2DOH} in astrophysical ice mixtures at \SI{10}{\kelvin} in the mid-infrared region.}
\label{tab:mixtures_CH2DOH_10K}
\setlength{\tabcolsep}{4pt}
\renewcommand{\arraystretch}{1.05}
\begin{tabular*}{\columnwidth}{@{\extracolsep{\fill}}lccc}
\toprule
Mode / \si{\rcm}\,(\si{\um}) & Binary mixture & \ch{H2O} matrix & \ch{CO} matrix \\
\midrule
\ch{C-D} stretch & 2179~(4.59) & 2183~(4.58) & 2181~(4.59) \\
\ch{CH2} wag   & 1326~(7.54) & 1329~(7.52) & 1325~(7.55) \\
\ch{CH2} twist & 1293~(7.73) & 1294~(7.73) & 1293~(7.73) \\
\ch{OCD} bend    & 919~(10.88) & 917~(10.90) & 920~(10.87) \\
\bottomrule
\end{tabular*}
\tablefoot{Band positions are given in \rcm, with corresponding wavelengths in \si{\um} shown in parentheses. Minor shifts of 1–4~\rcm indicate preservation of deuterated features across different ice matrices.}
\end{table}

\subsubsection{\ch{CHD2OH} mixtures}

For ice mixtures containing \ch{CHD2OH}, the transmission spectra show a set of well-defined deuteration-induced features that are preserved across all environments. A characteristic doublet is present in the binary mixture at \SI{1329}{\rcm} (\SI{7.52}{\um}, C--H bend) and \SI{1304}{\rcm} (\SI{7.67}{\um}, C--H bend). In the \ch{H2O}-rich ice, these bands shift slightly to \SI{1330}{\rcm} (\SI{7.52}{\um}) and \SI{1305}{\rcm} (\SI{7.66}{\um}), whereas in the \ch{CO}-rich ice they are observed at \SI{1329}{\rcm} (\SI{7.52}{\um}) and \SI{1304}{\rcm} (\SI{7.67}{\um}), respectively. The \ch{CD2} asymmetric stretch at \SI{2180}{\rcm} (\SI{4.59}{\um}) in the binary mixture shifts marginally to \SI{2181}{\rcm} (\SI{4.59}{\um}) in the \ch{H2O} matrix and remains at \SI{2180}{\rcm} (\SI{4.59}{\um}) in the \ch{CO} matrix.

\begin{figure}[htbp]
    \centering
    \includegraphics[scale=0.4]{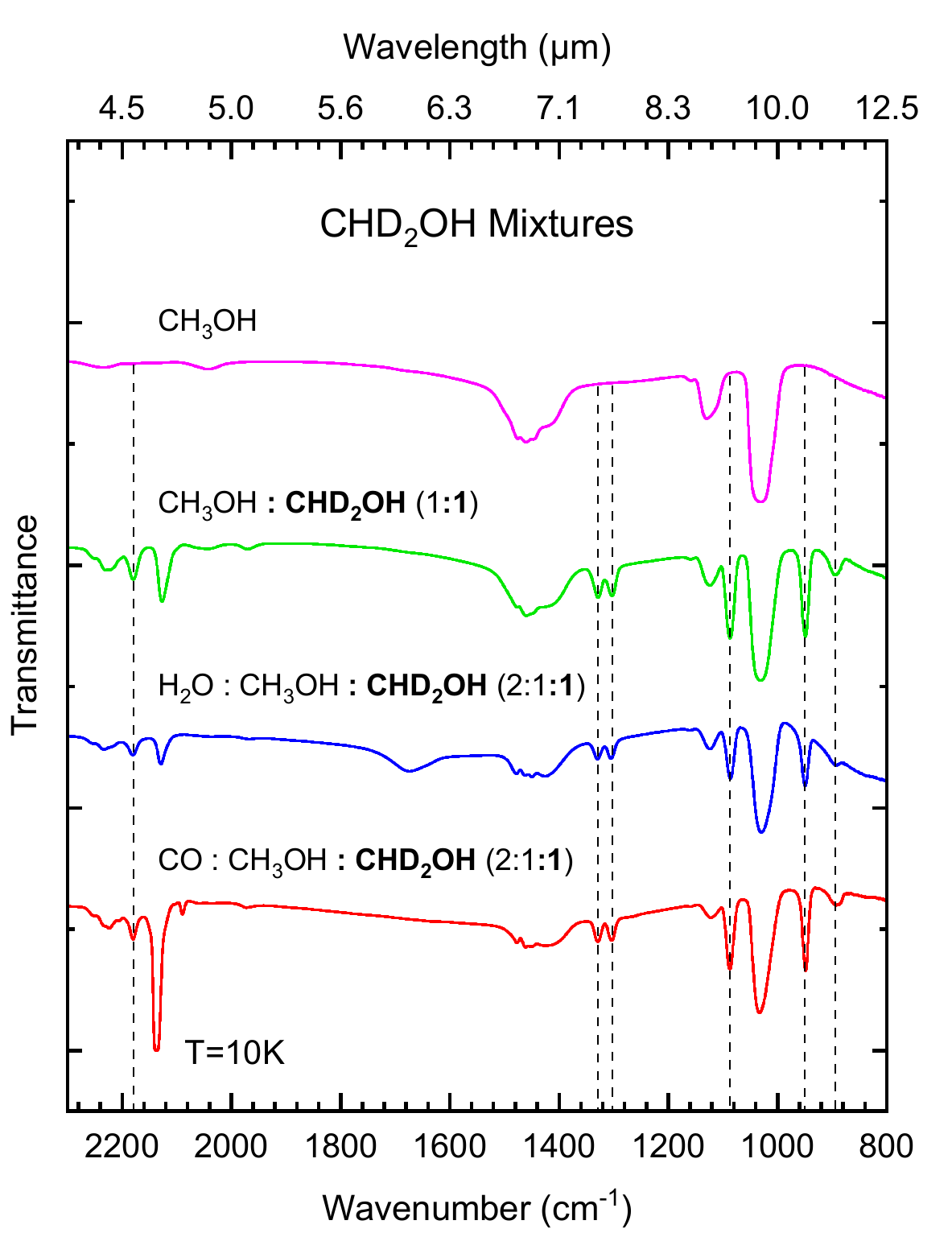}
        \caption{Transmission spectra of pure methanol ice and deuterated methanol (\ch{CHD2OH}) embedded in three astrophysical ice matrices at \SI{10}{\kelvin}. Key vibrational features associated with deuteration are marked by vertical dashed lines; their centres vary within 1--2~\rcm\ across the matrices (see Table~\ref{tab:mixtures_CHD2OH_10K}). Spectra are vertically offset for clarity.}
    \label{fig:mixtures_CHD2OH_10K}
\end{figure}

Lower-frequency \ch{CD2} modes are likewise conserved. The \ch{CD2} bend at \SI{1087}{\rcm} (\SI{9.20}{\um}) is unchanged in the binary and \ch{H2O} ices and shifts to \SI{1088}{\rcm} (\SI{9.19}{\um}) in the \ch{CO} ice. The \ch{CD2} wag at \SI{950}{\rcm} (\SI{10.53}{\um}) in the binary and \ch{H2O} ices moves to \SI{949}{\rcm} (\SI{10.54}{\um}) in the \ch{CO} ice, while the \ch{CD2} twist at \SI{894}{\rcm} (\SI{11.19}{\um}) in the binary ice appears at \SI{893}{\rcm} (\SI{11.20}{\um}) in both mixed matrices. All observed shifts are within \SIrange{1}{2}{\rcm}, consistent with the experimental uncertainty. Fig.~\ref{fig:mixtures_CHD2OH_10K} shows the transmission spectra of \ch{CHD2OH}-containing mixtures overlaid with that of pure methanol ice, illustrating deuteration-induced spectral changes; Table~\ref{tab:mixtures_CHD2OH_10K} summarises the frequencies of the key deuterated bands highlighted in the figure for each matrix.

\begin{table}[ht]
\caption{Deuterated vibrational modes of \ch{CHD2OH} in astrophysical ice mixtures at \SI{10}{\kelvin} in the mid-infrared region.}
\label{tab:mixtures_CHD2OH_10K}
\setlength{\tabcolsep}{4pt}
\renewcommand{\arraystretch}{1.05}
\begin{tabular*}{\columnwidth}{@{\extracolsep{\fill}}lccc}
\toprule
Mode / \si{\rcm}\,(\si{\um}) & Binary mixture & \ch{H2O} matrix & \ch{CO} matrix \\
\midrule
\ch{CD2} asym.\ stretch & 2180~(4.59) & 2181~(4.59) & 2180~(4.59) \\
\ch{C-H} bend             & 1329~(7.52) & 1330~(7.52) & 1329~(7.52) \\
\ch{C-H} bend           & 1304~(7.67) & 1305~(7.66) & 1304~(7.67) \\
\ch{CD2} bend           & 1087~(9.20) & 1087~(9.20) & 1088~(9.19) \\
\ch{CD2} wag            & 950~(10.53) & 950~(10.53) & 949~(10.54) \\
\ch{CD2} twist          & 894~(11.19) & 893~(11.20) & 893~(11.20) \\
\bottomrule
\end{tabular*}
\tablefoot{Band positions are given in \rcm, with corresponding wavelengths in \si{\um} shown in parentheses. Minor shifts of 1--2~\rcm indicate preservation of deuterated features across different ice matrices.}
\end{table}

\subsubsection{\ch{CD3OH} mixtures}

For ice mixtures containing \ch{CD3OH}, the transmission spectra reveal a consistent set of deuteration-induced features across the different environments. In \ch{CD3OH} ices, the C--O stretching band appears at \SI{982}{\rcm} (\SI{10.18}{\um}) in the binary mixture, shifting marginally to \SI{981}{\rcm} (\SI{10.19}{\um}) in the \ch{H2O}-rich ice and to \SI{983}{\rcm} (\SI{10.17}{\um}) in the \ch{CO}-rich ice. The \ch{CD3} symmetric stretching mode is observed at \SI{2072}{\rcm} (\SI{4.82}{\um}) in the binary mixture, moving to \SI{2075}{\rcm} (\SI{4.82}{\um}) in the \ch{H2O} matrix and to \SI{2074}{\rcm} (\SI{4.82}{\um}) in the \ch{CO} matrix.

\begin{figure}[htbp]
    \centering
    \includegraphics[scale=0.4]{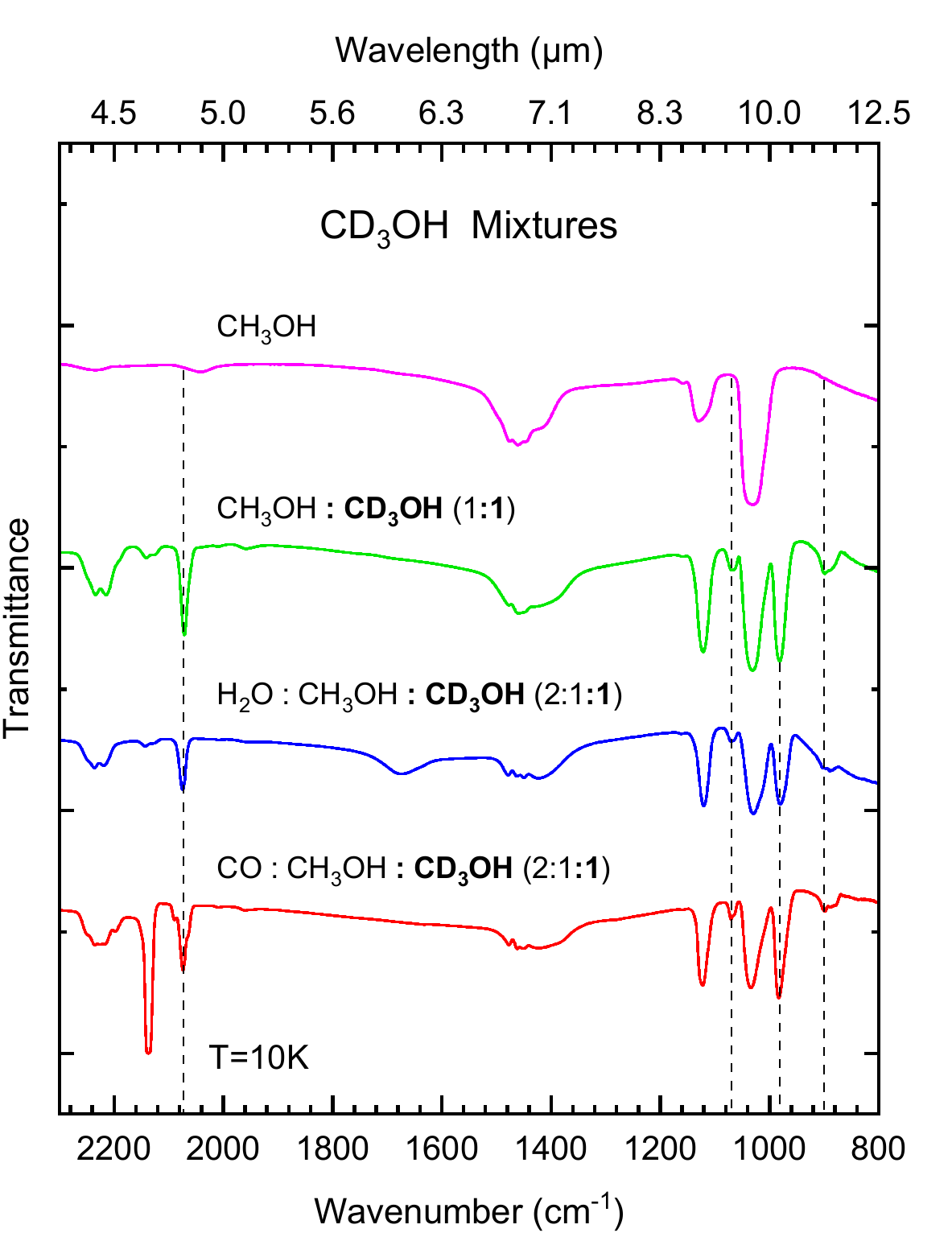}
    \caption{Transmission spectra of pure methanol ice and deuterated methanol (\ch{CD3OH}) embedded in three astrophysical ice matrices at \SI{10}{\kelvin}. Key vibrational features associated with deuteration are marked by vertical dashed lines; their centres vary within 1--3~\rcm\ across the matrices (see Table~\ref{tab:mixtures_CD3OH_10K}). Spectra are vertically offset for clarity.}
    \label{fig:mixtures_CD3OH_10K}
\end{figure}

The \ch{CD3} asymmetric bending mode occurs at \SI{1068}{\rcm} (\SI{9.36}{\um}) in the binary ice and shifts slightly to \SI{1069}{\rcm} (\SI{9.35}{\um}) in the \ch{H2O} mixture and to \SI{1070}{\rcm} (\SI{9.35}{\um}) in the \ch{CO} mixture. The \ch{CD3} rocking mode is located at \SI{899}{\rcm} (\SI{11.12}{\um}) in the binary ice, at \SI{903}{\rcm} (\SI{11.07}{\um}) in the \ch{H2O}-rich mixture, and again at \SI{899}{\rcm} (\SI{11.12}{\um}) in the \ch{CO}-rich ice. The observed variations in peak positions are limited to a few \si{\rcm} and therefore fall within the overall experimental uncertainty. Fig.~\ref{fig:mixtures_CD3OH_10K} shows the transmission spectra of \ch{CD3OH}-containing mixtures overlaid with that of pure methanol ice, enabling direct comparison of the spectral changes introduced by deuteration; Table~\ref{tab:mixtures_CD3OH_10K} summarises the frequencies of the key deuterated bands highlighted in the figure for each matrix.

\begin{table}[ht]
\caption{Deuterated vibrational modes of \ch{CD3OH} in astrophysical ice mixtures at \SI{10}{\kelvin} in the mid-infrared region.}
\label{tab:mixtures_CD3OH_10K}
\setlength{\tabcolsep}{4pt}
\renewcommand{\arraystretch}{1.05}
\begin{tabular*}{\columnwidth}{@{\extracolsep{\fill}}lccc}
\toprule
Mode / \si{\rcm}\,(\si{\um}) & Binary mixture & \ch{H2O} matrix & \ch{CO} matrix \\
\midrule
\ch{CD3} sym.\ stretch  & 2072~(4.82) & 2075~(4.82) & 2074~(4.82) \\
\ch{CD3} asym.\ bend     & 1068~(9.36) & 1069~(9.35) & 1070~(9.35) \\
\ch{C-O} stretch        & 982~(10.18) & 981~(10.19) & 983~(10.17) \\
\ch{CD3} rock     & 899~(11.12) & 903~(11.07) & 899~(11.12) \\
\bottomrule
\end{tabular*}
\tablefoot{Band positions are given in \rcm, with corresponding wavelengths in \si{\um} shown in parentheses. Minor shifts of 1--3~\rcm indicate preservation of deuterated features across different ice matrices.}
\end{table}

\subsubsection{\ch{CD3OD} mixtures}

For ice mixtures containing \ch{CD3OD}, the transmission spectra reveal a consistent set of deuteration-induced features across the different environments. In \ch{CD3OD}-containing ices, the overall spectral pattern closely resembles that observed for \ch{CD3OH}. The \ch{CD3} symmetric stretching mode appears at \SI{2072}{\rcm} (\SI{4.82}{\um}) in the binary mixture and shifts marginally to \SI{2074}{\rcm} (\SI{4.82}{\um}) in both the \ch{H2O}- and \ch{CO}-rich matrices. The \ch{CD3} asymmetric bending mode is observed at \SI{1066}{\rcm} (\SI{9.38}{\um}) in the binary ice, shifts to \SI{1068}{\rcm} (\SI{9.36}{\um}) in the \ch{H2O} matrix, and returns to \SI{1066}{\rcm} (\SI{9.38}{\um}) in the \ch{CO} matrix. The shifted C--O stretching band is located at \SI{979}{\rcm} (\SI{10.21}{\um}) in both the binary and \ch{H2O}-rich ices and at \SI{980}{\rcm} (\SI{10.20}{\um}) in the \ch{CO}-rich ice.

Lower-frequency modes are likewise stable across all environments. The \ch{CD3} rocking mode remains at \SI{899}{\rcm} (\SI{11.12}{\um}) in all mixtures, while a second \ch{CD3} rocking mode is observed at \SI{836}{\rcm} (\SI{11.96}{\um}) in the binary and \ch{H2O}-rich ices and at \SI{834}{\rcm} (\SI{11.99}{\um}) in the \ch{CO}-rich ice. All observed shifts are at the level of a few \si{\rcm} and fall within the overall experimental uncertainty.

A weak feature at \SI{1234}{\rcm} (\SI{8.10}{\um}) is present in spectra of both pure and mixed \ch{CD3OD} ices. This feature is attributed to trace contamination in the original \ch{CD3OD} sample and is therefore excluded from the band assignments discussed here.

\begin{figure}[htbp]
    \centering
    \includegraphics[scale=0.4]{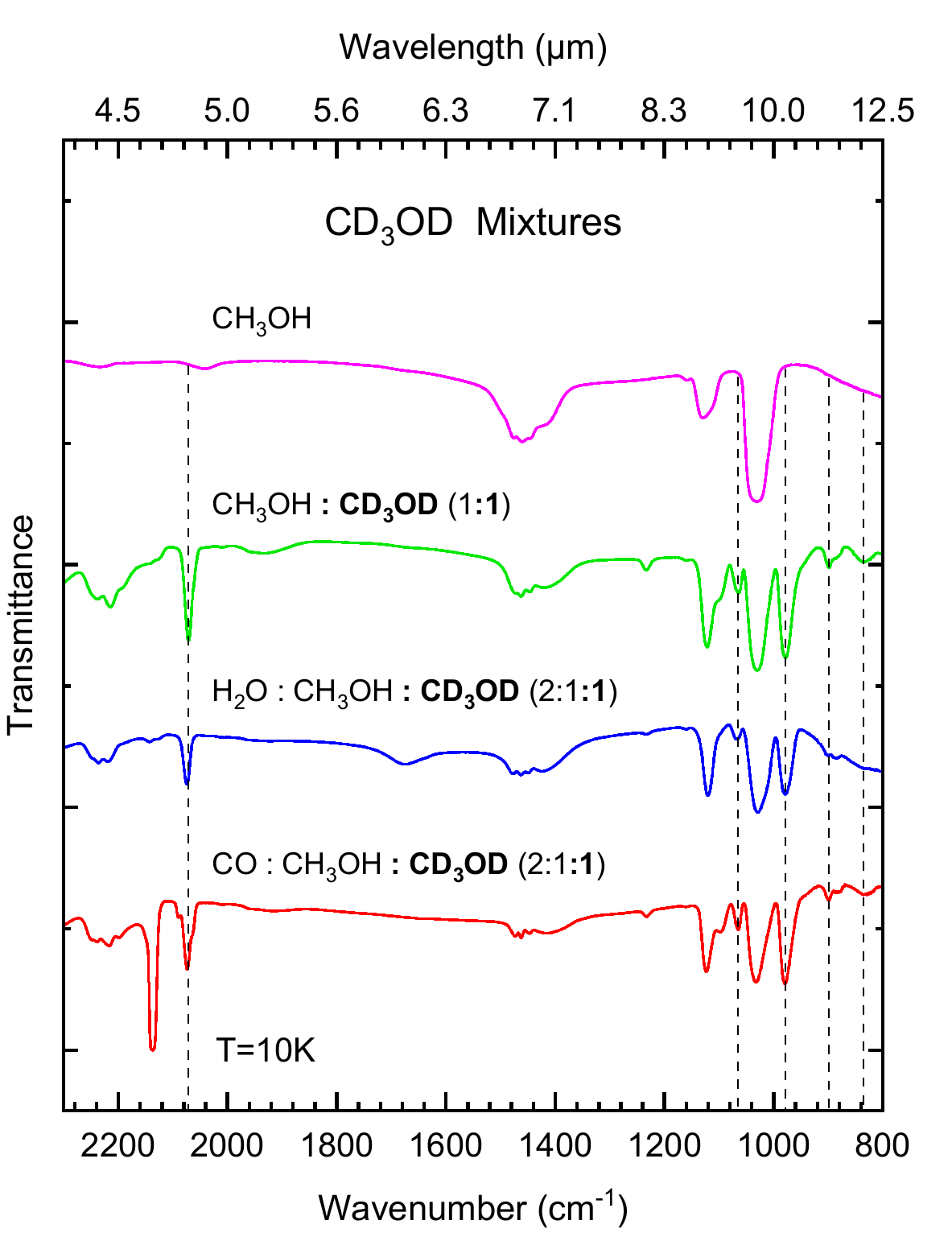}
    \caption{Transmission spectra of pure methanol ice and deuterated methanol (\ch{CD3OD}) embedded in three astrophysical ice matrices at \SI{10}{\kelvin}. Key vibrational features associated with deuteration are marked by vertical dashed lines; their centres vary within 1--2~\rcm\ across the matrices (see Table~\ref{tab:mixtures_CD3OD_10K}). Spectra are vertically offset for clarity.}
\label{fig:mixtures_CD3OD_10K}
\end{figure}

Overall, the characteristic deuteration-induced spectral features of \ch{CD3OD} appear to be preserved across the different astrophysical ice mixtures, with only minor variations comparable to the experimental uncertainty. Fig.~\ref{fig:mixtures_CD3OD_10K} shows the transmission spectra of \ch{CD3OD}-containing mixtures overlaid with that of pure methanol ice, enabling direct comparison of the spectral features associated with deuteration; Table~\ref{tab:mixtures_CD3OD_10K} summarises the frequencies of the key deuterated bands highlighted in the figure for each matrix.

\begin{table}[ht]
\caption{Deuterated vibrational modes of \ch{CD3OD} in astrophysical ice mixtures at \SI{10}{\kelvin} in the mid-infrared region.}
\label{tab:mixtures_CD3OD_10K}
\setlength{\tabcolsep}{4pt}
\renewcommand{\arraystretch}{1.05}
\begin{tabular*}{\columnwidth}{@{\extracolsep{\fill}}lccc}
\toprule
Mode / \si{\rcm}\,(\si{\um}) & Binary mixture & \ch{H2O} matrix & \ch{CO} matrix \\
\midrule
\ch{CD3} sym.\ stretch   & 2072~(4.82) & 2074~(4.82) & 2074~(4.82) \\
\ch{CD3} asym.\ bend     & 1066~(9.38) & 1068~(9.36) & 1066~(9.38) \\
\ch{C-O} stretch         & 979~(10.21) & 979~(10.21) & 980~(10.20) \\
\ch{CD3} rock      & 899~(11.12) & 899~(11.12) & 899~(11.12) \\
\ch{CD3} rock           & 836~(11.96) & 836~(11.96) & 834~(11.99) \\
\bottomrule
\end{tabular*}
\tablefoot{Band positions are given in \rcm, with corresponding wavelengths in \si{\um} shown in parentheses. Minor shifts of 1--2~\rcm indicate preservation of deuterated features across different ice matrices.}
\end{table}

\section{Discussion}
\label{sec:discussion}

Deuterated methanol is an exceptionally sensitive probe of grain-surface chemistry in the cold interstellar medium. While the elemental \mbox{D/H} ratio in the local ISM is $(2.0\pm0.1)\times10^{-5}$ \citep{VidalMadjar2002, Linsky2006, Prodanovic2010}, the molecular D/H ratio in methanol can be enhanced by several orders of magnitude in star-forming regions. Towards the Class 0 protostar IRAS\,16293$-$2422, the abundance ratio $[\ch{CH2DOH}]/[\ch{CH3OH}]$ reaches $0.9\pm0.3$ \citep{Parise2002}, and even Solar System material such as comet 67P/Churyumov--Gerasimenko shows a significantly enhanced total monodeuterated methanol fraction, $([\ch{CH2DOH}]+[\ch{CH3OD}]) / [\ch{CH3OH}] = 0.055 \pm 0.005$, corresponding to an upper limit estimate \citep{Drozdovskaya2020}. Singly, doubly, and triply deuterated isotopologues (\ch{CH2DOH}, \ch{CHD2OH}, \ch{CD3OH}) have been detected in prestellar cores and hot corino sources \citep{Parise2004, Parise2006, Bizzocchi2014, Bianchi2017, Manigand2019, Lin2023, Spezzano2025}, confirming that deuterium is incorporated efficiently into methanol ices at $10$–$20$\,K and that methanol deuteration predominantly traces cold prestellar conditions \citep{Ceccarelli2014, Caselli2012b, Bianchi2017}.

Methanol itself is formed almost exclusively on dust grains through successive H-atom additions to CO ice \citep{Watanabe2002, Fuchs2009}. This hydrogenation network is among the best constrained in astrochemistry thanks to extensive laboratory and theoretical work \citep{Osamura2004, Nagaoka2005, Hidaka2009, Chuang2015}. Gas–grain models reproduce the large abundances of \ch{CH2DOH}, \ch{CHD2OH}, and \ch{CD3OH} once the atomic D/H ratio in the gas phase climbs to $\sim$0.1–0.2 after catastrophic CO freeze-out \citep{Taquet2012, Caselli2022}. By contrast, the hydroxyl-deuterated isotopologue \ch{CH3OD} is systematically underabundant relative to \ch{CH2DOH} in protostellar sources and hot corinos \citep{Parise2006, Jorgensen2018, Manigand2019}. Proposed explanations include (i) kinetic and energetic preferences for H abstraction and addition during surface hydrogenation \citep{Nagaoka2005}; (ii) post-desorption destruction of \ch{CH3OD} through gas-phase proton-transfer cycles \citep{Charnley1997, Sipila2015}; and (iii) selective H/D exchange with \ch{H2O} or \ch{NH3} ices that removes D from the hydroxyl group \citep{Kawanowa2004, Ratajczak2009, Lamberts2015, Faure2015}. Because the hydroxyl group readily participates in hydrogen bonding, such exchange reactions may partially erase the original deuteration signature at the OH site, complicating direct links between \ch{CH3OD} abundances and formation pathways. The persistent \ch{CH3OD} deficit therefore offers an additional constraint on the physical history and post-formation processing of interstellar ices.

Laboratory experiments and theoretical studies further indicate that the dominant formation pathways of singly and doubly deuterated methanol differ in their sensitivity to the timing of deuterium availability on grain surfaces. Monodeuterated methanol (\ch{CH2DOH}) is thought to form predominantly through H–D substitution reactions in pre-existing methanol, following abstraction of an H atom from the methyl group and subsequent D addition. This mechanism is efficient at $\sim$10–20~K and strongly favours methyl deuteration over hydroxyl substitution, owing to both kinetic barriers and reaction energetics \citep{Osamura2004, Nagaoka2005, Nagaoka2007, Goumans2011}. In contrast, doubly deuterated methanol (\ch{CHD2OH}) is more readily produced earlier in the CO hydrogenation sequence, via H–D substitution reactions in formaldehyde (\ch{HDCO}, \ch{D2CO}) followed by further hydrogenation to methanol \citep{Hidaka2009, Chuang2015}. These distinct pathways imply that the relative abundances of \ch{CH2DOH} and \ch{CHD2OH} reflect not only the atomic D/H ratio in the accreting gas, but also the temporal overlap between CO hydrogenation and deuterium enrichment on grain surfaces \citep{Aikawa2012, Taquet2014}. The simultaneous presence of both isotopologues therefore supports scenarios in which hydrogenation and deuteration proceed simultaneously during the cold prestellar phase, rather than in strictly sequential stages, in agreement with laboratory constraints and gas–grain chemical models \citep{Taquet2012, Ceccarelli2014, Drozdovskaya2020}.

Beyond serving as a diagnostic tool, methanol is a key precursor to complex organic molecules (COMs). UV or thermal processing of \ch{CH3OH} ices liberates radicals (e.g. \ch{CH3}, \ch{CH3O}, \ch{CH2OH}) that drive COM formation \citep{Oberg2009, Bertin2016}. Laboratory studies demonstrate that deuterium initially stored in \ch{CH2DOH} or \ch{CH3OD} can be transferred to COMs, and that the efficiency depends on whether the donor is the methyl or the hydroxyl site \citep{Oba2017}. Since methanol deuteration is largely established during the cold prestellar phase and is expected to be halted above $\sim20$\,K, the isotopic composition of methanol provides a fossil record of early ice chemistry that may be partially inherited by more complex species. An accurate inventory of deuterium among methanol isotopologues is therefore essential for interpreting ongoing JWST and ALMA surveys of deuterated COMs.

High-resolution infrared spectra obtained in our experiments for amorphous methanol ices at 10~K and for partially crystalline ices after annealing to 120~K reveal several stable diagnostic features across all methanol isotopologues, studied both in pure form and in \ch{H2O}- and \ch{CO}-rich matrices representative of interstellar ice environments. A key spectroscopic signature is the characteristic doublet observed in the bending region of singly and doubly deuterated methanol: \ch{CH2DOH} exhibits a well-separated pair at 1326 and 1293~\rcm (7.54 and 7.73~\si{\um}), while \ch{CHD2OH} shows a corresponding double feature at 1329 and 1301~\rcm (7.52 and 7.69~\si{\um}). These splittings persist across different ice matrices and temperatures, with only minor shifts at the level of the experimental uncertainty, supporting their use as reliable spectral fingerprints for identifying methanol isotopologues in astronomical ice spectra and for tracing prestellar ice compositions into later evolutionary stages.

Our measurements span two representative temperature regimes: a low-temperature regime at 10~K and a high-temperature regime after annealing to 120~K. The 10~K spectra probe conditions characteristic of cold interstellar ices in dense molecular clouds and prestellar cores, where methanol and its isotopologues are thought to form on grain surfaces and where methanol is predominantly observed in the solid phase. In addition to these endpoint measurements, a limited set of intermediate-temperature annealing experiments was performed between 10 and 120~K, revealing only modest band shifts below the crystallisation temperature. The most pronounced spectral changes occur upon annealing to 120~K, where methanol undergoes a pronounced phase transition from amorphous to crystalline ice, leading to systematic frequency shifts and band sharpening that dominate the spectral evolution. Such temperatures are relevant to later stages of star formation, when interstellar ices experience substantial thermal processing in warm protostellar environments prior to sublimation. While the present data provide robust reference spectra at these two temperature limits, extrapolation to intermediate temperatures should be treated with caution, particularly during gradual warm-up phases where partial ice restructuring, segregation, or isotope exchange processes may occur.

To facilitate comparison with astronomical observations, the experimental mixing ratios were chosen to reproduce the maximum observed \ch{CH2DOH} abundance ([\ch{CH2DOH}]/[\ch{CH3OH}] $\sim$ 1) and a lower limit set by experimental sensitivity ([\ch{CH2DOH}]/[\ch{CH3OH}] $\sim$ 0.1). In addition, theoretical models by \citet{Riedel2023} suggest that deuteration can be efficient enough for singly deuterated methanol to exceed its parent species in abundance. We therefore performed a third set of experiments exploring the case [\ch{CH2DOH}]/[\ch{CH3OH}] $>$ 1. All other isotopologues were assigned the same mixing ratio as \ch{CH2DOH} in these experiments.

\section{Conclusions}

We have presented a systematic laboratory investigation of the infrared spectra of solid methanol and its five main isotopologues. Reliable band positions and tentative vibrational assignments across the near-, mid-, and far-infrared are established for the pure methanol isotopologues, supported by complementary anharmonic vibrational calculations and additional spectra recorded after annealing to \SI{120}{\kelvin}. The behaviour of deuteration-induced bands in the mid-infrared is examined in binary, \ch{CO}-rich, and \ch{H2O}-rich ice mixtures at \SI{10}{\kelvin}. For all isotopologues, the deuteration-induced bands are largely preserved, with only minor matrix-dependent shifts that remain within the experimental uncertainty. The resulting dataset comprises a coherent laboratory reference for the identification of deuterated methanol isotopologues in interstellar ices and supports the interpretation of current and forthcoming infrared observations, including those from JWST. A quantitative determination of integrated band strengths and their dependence on temperature and ice composition remains an important objective for future work.

\section*{Data availability}
All laboratory spectra acquired during this study are openly available at \url{https://doi.org/10.5281/zenodo.18257626}.

\begin{acknowledgements}
This work was supported by the Max Planck Society.
We thank Christian Deysenroth and his team for designing and developing the experimental setup and for their continuous assistance in laboratory development. This article is based on work from the COST Action CA20129 – Multiscale Irradiation and Chemistry Driven Processes and Related Technologies (MultIChem) and COST Action CA22133 – The Birth of Solar Systems (PLANETS), supported by COST (European Cooperation in Science and Technology).
\end{acknowledgements}

\bibliographystyle{aa_url}
\bibliography{references}

\onecolumn
\appendix

\section{Mid-infrared bands of methanol isotopologues}
\label{app:midIR_bands}

The vibrational bands of solid methanol and its isotopologues in the
\SIrange{2300}{800}{\rcm} (\SIrange{4.35}{12.5}{\um}) mid-infrared region are 
summarised here. This spectral range encompasses C--D stretching modes, 
\ch{CH3} deformation modes, O--H/O--D bending vibrations, rocking motions, 
as well as twisting and wagging modes, together with the C--O stretching band. 
Collectively, these features provide some of the most diagnostically relevant 
spectral signatures of methanol-bearing interstellar ices. The bands compiled in 
Table~\ref{tab:midIR_bands} are extracted from the laboratory dataset presented 
in the main text and are intended as a concise, observation-oriented reference. 
Notable examples include the characteristic doublet of \ch{CH2DOH} at 1293 and 
1326~\rcm\ (7.73 and 7.54~\si{\um}) and the analogous doublet of \ch{CHD2OH} 
at 1301 and 1329~\rcm\ (7.69 and 7.52~\si{\um}). In astronomical spectra, these
features may be affected by continuum placement, silicate absorption in the
$\sim$10~\si{\um} region, and potential overlap with the \ch{CH4} band
near 7.7~\si{\um}. Their identification may therefore require careful
analysis, particularly as a function of ice composition and relative molecular abundances. All band positions reported here correspond to amorphous ices deposited at \SI{10}{\kelvin}. Band assignments remain tentative in spectral regions where
multiple bending and deformation modes overlap. A more complete overview of the
vibrational modes observed in this study, including the stretching and torsional
regions, is provided in the tables and figures presented in
Sect.~\ref{sec:results}.

\begin{table*}[h]
\centering
\caption{Mid-infrared vibrational modes of amorphous methanol ice (\ch{CH3OH}) and its isotopologues between 2300 and 800~\rcm\ at \SI{10}{\kelvin}.}
\label{tab:midIR_bands}
\footnotesize
\setlength{\tabcolsep}{6pt}
\renewcommand{\arraystretch}{1.05}
\begin{tabular*}{\textwidth}{@{\extracolsep{\fill}}lcccccc}
\toprule
Mode / \si{\rcm}\,(\si{\um}) & \ch{CH3OH} & \ch{CH3OD} & \ch{CH2DOH} & \ch{CHD2OH} & \ch{CD3OH} & \ch{CD3OD} \\
\midrule
\ch{CD3} asym.\ stretch & --- & --- & --- & --- & 2233~(4.48) & 2244~(4.46) \\
\ch{CD3} asym.\ stretch & --- & --- & --- & --- & 2215~(4.51) & 2216~(4.51) \\
\ch{CD2} asym.\ stretch & --- & --- & --- & 2222~(4.50) & --- & --- \\
\ch{CD2} asym.\ stretch & --- & --- & --- & 2180~(4.59) & --- & --- \\
\ch{C-D} stretch        & --- & --- & 2242~(4.46) & --- & --- & --- \\
\ch{C-D} stretch        & --- & --- & 2179~(4.59) & --- & --- & --- \\
\ch{C-D} stretch        & --- & --- & 2146~(4.66) & --- & --- & --- \\
\ch{CD2} sym.\ stretch  & --- & --- & --- & 2126~(4.70) & --- & --- \\
\ch{CD3} sym.\ stretch  & --- & --- & --- & --- & 2071~(4.83) & 2071~(4.83) \\
\ch{CH3} deformation    & 1461~(6.84) & --- & --- & --- & --- & --- \\
\ch{CH3} asym.\ bend    & --- & 1470~(6.80) & --- & --- & --- & --- \\
\ch{CH3} sym.\ bend     & --- & 1445~(6.92) & --- & --- & --- & --- \\
\ch{CH2} bend           & --- & --- & 1464~(6.83) & --- & --- & --- \\
\ch{C-H} bend           & --- & --- & --- & 1329~(7.52) & --- & --- \\
\ch{C-H} bend           & --- & --- & --- & 1301~(7.69) & --- & --- \\
\ch{O-H}/\ch{O-D} bend  & 1461~(6.84) & 942~(10.62) & 1464~(6.83) & 1424~(7.02) & 1398~(7.15) & 1123~(8.90) \\
\ch{CH3} rock           & --- & 1234~(8.10) & --- & --- & --- & --- \\
\ch{CH3} rock           & 1158~(8.64) & 1161~(8.61) & --- & --- & --- & --- \\
\ch{CH3} rock           & 1129~(8.86) & 1123~(8.90) & --- & --- & --- & --- \\
\ch{CD3} sym.\ bend     & --- & --- & --- & --- & 1122~(8.91) & 1123~(8.90) \\
\ch{CH2} wag            & --- & --- & 1326~(7.54) & --- & --- & --- \\
\ch{CH2} twist          & --- & --- & 1293~(7.73) & --- & --- & --- \\
\ch{CH2} rock           & --- & --- & 1108~(9.03) & --- & --- & --- \\
\ch{CD2} bend           & --- & --- & --- & 1088~(9.19) & --- & --- \\
\ch{CH2} rock           & --- & --- & 1087~(9.20) & --- & --- & --- \\
\ch{CD3} asym.\ bend    & --- & --- & --- & --- & 1067~(9.37) & 1062~(9.42) \\
C--O stretch            & 1031~(9.70) & 1031~(9.70) & 1036~(9.65) & 1031~(9.70) & 981~(10.19) & 975~(10.26) \\
\ch{CD2} wag            & --- & --- & --- & 949~(10.54) & --- & --- \\
\ch{OCD} bend           & --- & --- & 919~(10.88) & --- & --- & --- \\
\ch{CD2} twist          & --- & --- & --- & 895~(11.17) & --- & --- \\
\ch{CD3} rock           & --- & --- & --- & --- & 898~(11.14) & 900~(11.11) \\
\ch{CD3} rock           & --- & --- & --- & --- & --- & 832~(12.02) \\
\bottomrule
\end{tabular*}
\tablefoot{Band positions are given in \rcm, with corresponding wavelengths in \si{\um} shown in parentheses. ``---'' indicates that the mode is not observed or not applicable for that isotopologue. Assignments are tentative.}
\end{table*}

\section{Pure ice isotopologue spectra at 120~K}
\label{app:spectra_120K}

\begin{figure*}[t]
  \centering
  \includegraphics[width=0.95\textwidth]{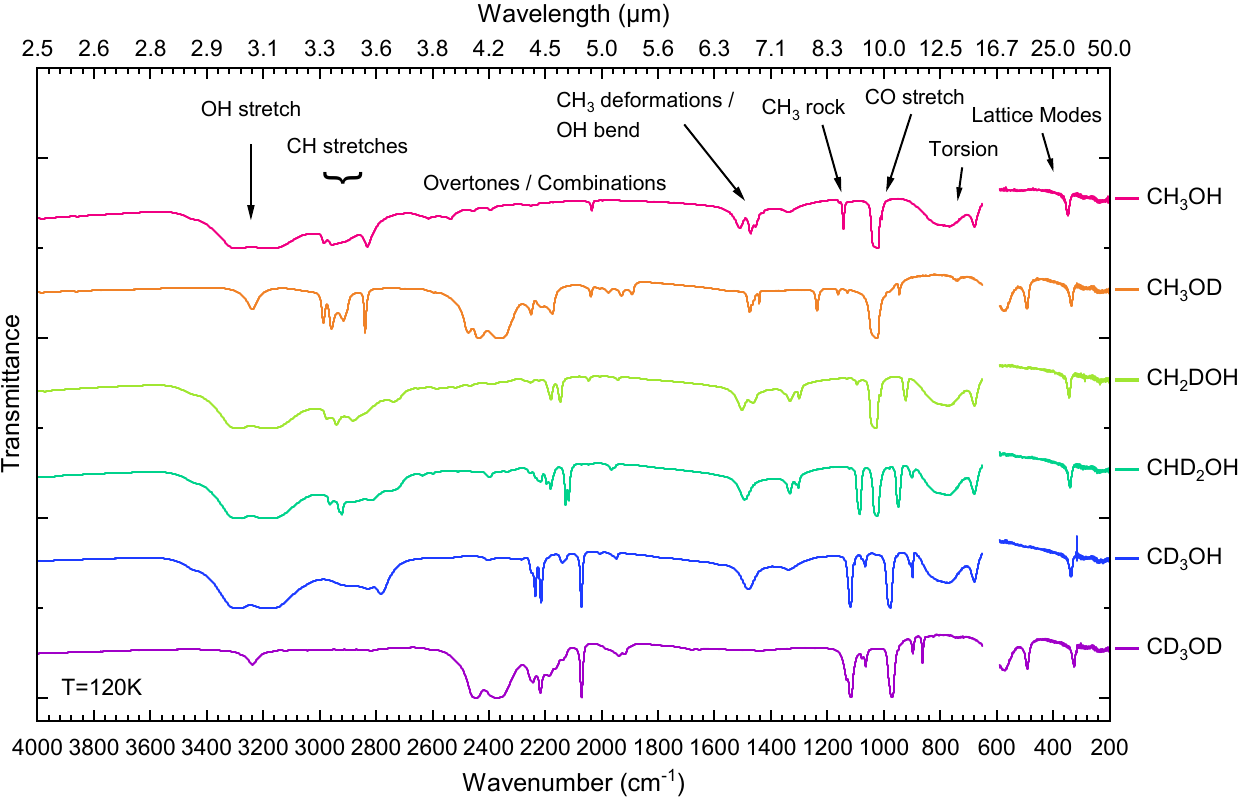}
    \caption{Transmission spectra of solid methanol and its five isotopologues at 120 K. The samples are identical to those shown in Fig.~\ref{fig:CH3OH_isotopologues_10K}. The region between \SIrange{650}{590}{\rcm} (\SIrange{15.4}{16.9}{\um}), affected by increased noise from the beamsplitter, has been removed. The main \ch{CH3OH} vibrational modes are indicated at the top. Spectra are vertically offset for clarity.}
  \label{fig:CH3OH_isotopologues_120K}
\end{figure*}

We present transmission spectra of pure methanol ice (\ch{CH3OH}) and its five isotopologues
obtained from the same samples shown in Fig.~\ref{fig:CH3OH_isotopologues_10K} after warming up from 10~K to 120~K at a rate of 1-2~K\,min\(^{-1}\). Compared to the 10~K spectra, the 120~K data (Fig.~\ref{fig:CH3OH_isotopologues_120K}) show pronounced band sharpening, the appearance of band splittings, and small shifts in peak positions, typically
limited to 5--10~\si{\rcm}, across the principal vibrational modes. These spectral
changes are consistent with crystallisation of the ice. Despite the
structural reorganisation, isotopic signatures remain clearly identifiable,
particularly for modes associated with C--D and O--D vibrations. Crystallisation
also resolves the previously blended O--H bending and \ch{CH3} deformation
region, thereby facilitating the interpretation and tentative assignment of the
corresponding bands observed at 10~K. In the far-infrared, further band 
sharpening is observed, together with a systematic red--shift upon deuteration 
of the hydroxyl group. Annealing to \SI{120}{\kelvin} therefore provides a 
laboratory analogue of the thermal processing experienced by icy grain mantles 
as protostellar systems evolve from the cold prestellar phase toward warmer 
inner regions. The availability of both low-temperature (10~K) and annealed 
(120~K) spectra provides complementary benchmarks for the interpretation of 
infrared observations of interstellar ices. The 120~K spectra presented here 
are primarily intended to support temperature-dependent studies and are not 
analysed further in the present work. Finally, we note that the 120~K spectra 
of \ch{CH3OD} and \ch{CD3OD} exhibit a weak residual feature in the O--H 
stretching region between 3200 and 3600~\si{\rcm} (2.78--3.13~\si{\um}). As 
these isotopologues do not contain an O--H group, this feature is attributed 
to trace contamination, most likely arising from small amounts of undeuterated 
methanol or water introduced during sample preparation. While this artefact does not affect the qualitative conclusions of this work, it should be considered in any quantitative analysis of these spectra.

\section{Concentration-dependent spectra of methanol mixtures}
\label{app:mixture_series}

\begin{figure*}[t]
  \centering
  \includegraphics[width=0.95\textwidth]{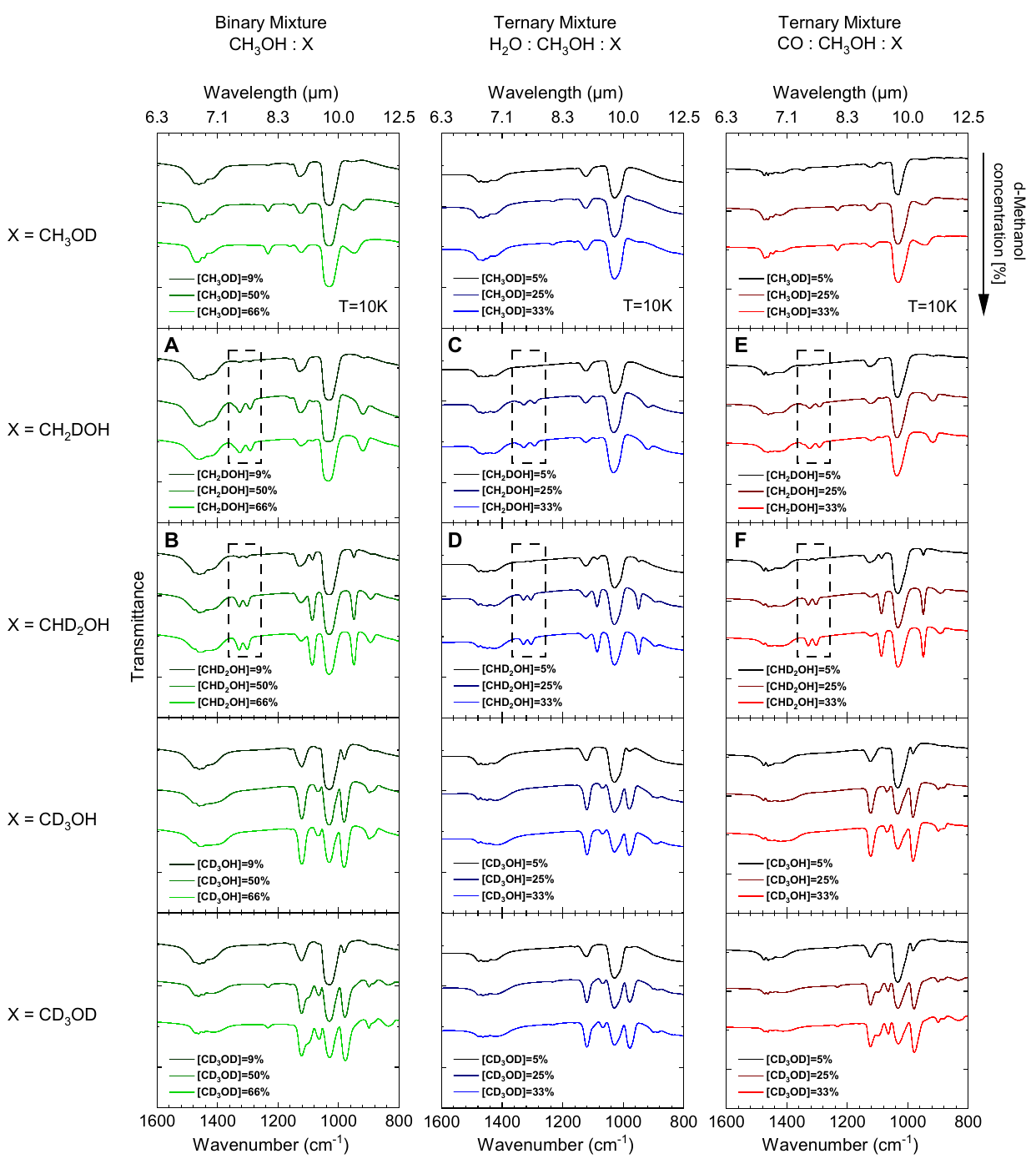}
  \caption{Collection of mid-infrared transmission spectra in the
    \SIrange{1600}{800}{\rcm}
    (\SIrange{6.25}{12.5}{\um}) region of binary \ch{CH3OH}\,:\,$X$ ices (left panel)
    and ternary \ch{H2O}\,:\,\ch{CH3OH}\,:\,$X$ (centre panel) and
    \ch{CO}\,:\,\ch{CH3OH}\,:\,$X$ (right panel) ices, where
    $X \in \{\ch{CH3OD}, \ch{CH2DOH}, \ch{CHD2OH}, \ch{CD3OH}, \ch{CD3OD}\}$.
    Each panel presents spectra acquired at three increasing concentrations of $X$
    (9\,\%, 50\,\%, and 66\,\% for binary mixtures; 5\,\%, 25\,\%, and 33\,\% for
    ternary mixtures). Highlighted spectral regions (A–F) indicate the characteristic
    double-peak features of \ch{CH2DOH} and \ch{CHD2OH} in different ice
    environments, analysed further in Fig.~\ref{fig:integrated_bands}. All spectra
    were recorded at \SI{10}{\kelvin} and are vertically offset for clarity.
    Additional experimental details are provided in
    Appendix~\ref{app:mixture_series}.}
  \label{fig:mixture_series}
\end{figure*}

\begin{figure*}[t]
    \centering
    \includegraphics[width=0.65\textwidth]{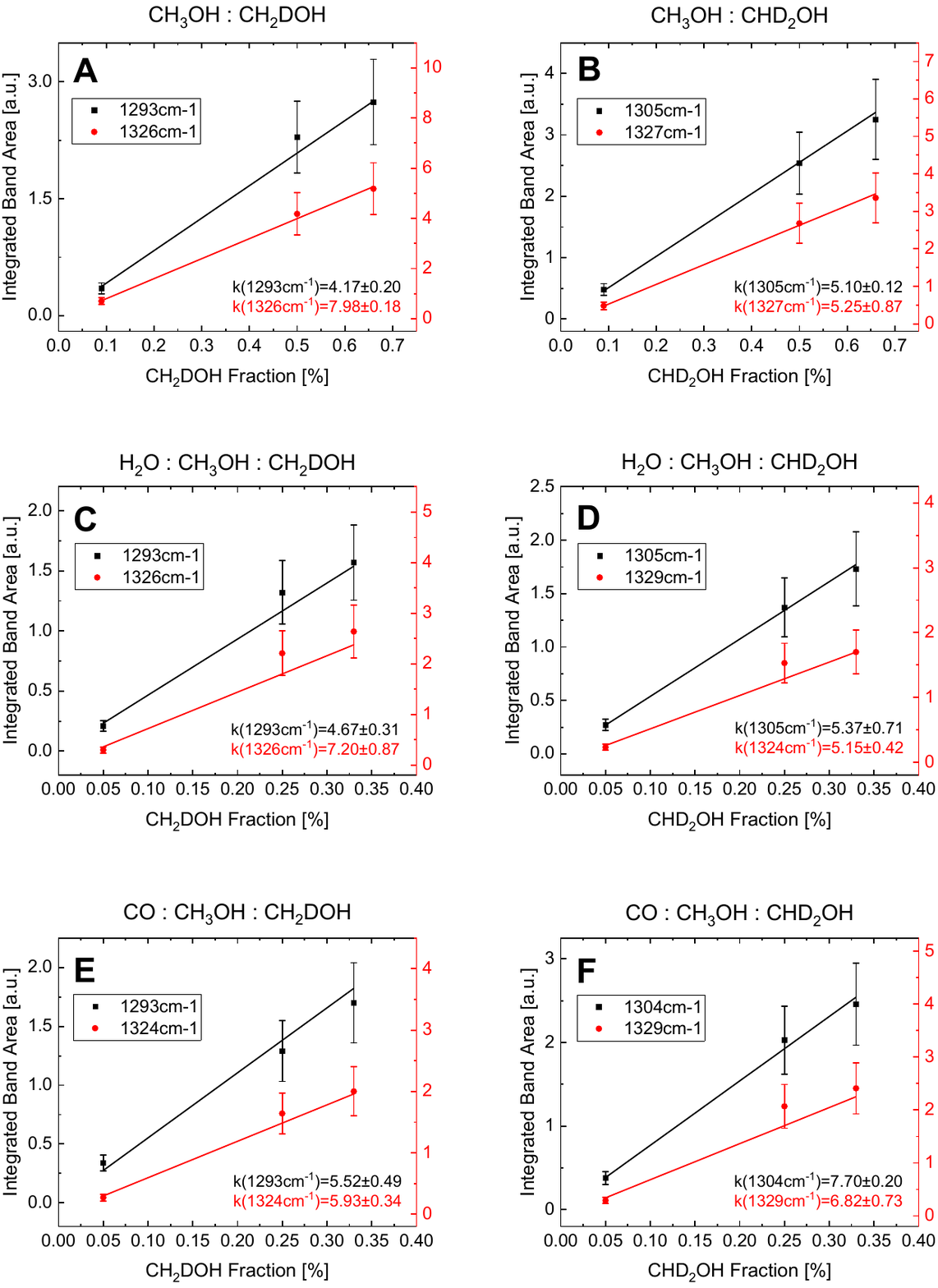}
    \caption{Integrated areas of the doublets marked in regions A–F of Fig.~\ref{fig:mixture_series}. Panels A–F correspond to the same spectra shown in Fig.~\ref{fig:mixture_series}. For each spectral window, the two peak components were integrated separately as a function of increasing deuterated methanol concentration, and the resulting band areas were plotted against the total fraction of \ch{CH2DOH} and \ch{CHD2OH} in the ice, showing an approximately linear dependence across all mixture types.}
    \label{fig:integrated_bands}
\end{figure*}

Fig.~\ref{fig:mixture_series} presents mid-infrared transmission spectra between \SIrange{1600}{800}{\rcm} (\SIrange{6.25}{12.5}{\um}) illustrating the behaviour of partially deuterated methanol isotopologues at
increasing concentrations in binary and ternary laboratory ice analogues. For each
isotopologue $X \in \{\ch{CH3OD}, \ch{CH2DOH}, \ch{CHD2OH}, \ch{CD3OH}, \ch{CD3OD}\}$,
we prepared a binary \ch{CH3OH}\,:\,$X$ mixture (left panel) with
\SIlist{9;50;66}{\percent} of $X$, as well as two ternary mixtures,
\ch{H2O}\,:\,\ch{CH3OH}\,:\,$X$ (centre panel) and
\ch{CO}\,:\,\ch{CH3OH}\,:\,$X$ (right panel), with
\SIlist{5;25;33}{\percent} of $X$. Percentages refer to the fraction of deuterated methanol relative to the total ice composition. All films were deposited \emph{in situ} onto a silicon substrate held at
\SI{10}{\kelvin} under $P \approx 10^{-7}$~mbar and recorded using a Bruker
Vertex~70v FTIR spectrometer at a spectral resolution of
\SI{1}{\rcm}. Successive spectra within each panel are vertically offset for
clarity. With increasing deuterated methanol concentration, C--D and O--D absorption features systematically increase in intensity, accompanied by a corresponding
reduction of neighbouring C--H and O--H bands. A particularly distinctive behaviour is observed for \ch{CH2DOH} and
\ch{CHD2OH}, both of which exhibit double-peak structures in the
\SIrange{1350}{1250}{\rcm} range. For \ch{CH2DOH}, this doublet appears near
1293 and 1326~\si{\rcm} (7.73 and 7.54~\si{\um}), while for
\ch{CHD2OH} it is found near 1301 and 1329~\si{\rcm} (7.69 and
7.52~\si{\um}). The six spectra in which these features are explicitly
labelled A–F in Fig.~\ref{fig:mixture_series} correspond to all three ice matrices
for both isotopologues. Fig.~\ref{fig:integrated_bands} shows the integrated band areas of these
double-peak regions as a function of the total isotopologue concentration. In all
matrices, the resulting trends are approximately linear, indicating that these band
pairs provide a quantitative tracer of \ch{CH2DOH} and \ch{CHD2OH} even in mixed
ices. Together, this concentration-dependent spectral set provides a laboratory
reference for the identification of partially deuterated
methanol in water- and CO-rich interstellar ice analogues.

\clearpage

\section{Mid- and near-infrared overtone and combination bands}
\label{app:NIR_bands}

Table~\ref{tab:nir_comparison} compares selected near-infrared
overtone and combination band positions measured in solid methanol
isotopologues at \SI{10}{\kelvin} with experimental and anharmonic
VPT2 frequencies reported by \citet{Grabska2017} for dilute
\ch{CCl4} solutions. The mean absolute deviation between our
measurements and the closest reported band positions is approximately
\SI{11}{\rcm}, with individual deviations ranging from
\SIrange{2}{32}{\rcm}. For the majority of selected bands, the solid-state values tend to lie closer to calculated anharmonic frequencies than to solution-phase experimental values. Table~\ref{tab:overtones_combinations} summarises the mid-infrared
overtone and combination bands observed in solid methanol and its isotopologues at \SI{10}{\kelvin}. The listed assignments
are guided by the anharmonic vibrational calculations presented in
this work and indicate the dominant contributing transition for
each observed feature. Owing to strong anharmonicity, lattice
effects, and extensive mode mixing in this spectral region, several
bands are expected to contain additional weaker contributions
beyond the dominant transition identified here.

\begin{table*}[h]
\centering
\caption{Near-infrared combination bands of solid methanol and selected isotopologues at \SI{10}{\kelvin}, compared with experimental values measured in dilute \ch{CCl4} solution and anharmonic calculated frequencies from \citet{Grabska2017}.}
\label{tab:nir_comparison}
\footnotesize
\setlength{\tabcolsep}{4pt}
\renewcommand{\arraystretch}{1.0}
\begin{tabular}{lccccl}
\toprule
Isotopologue & This work ($\nu$) & This work ($\lambda$) & \citet{Grabska2017}$^a$ & $\Delta$$^b$ & Tentative assignment \\
             & [\rcm]    & [\um]                 & [\rcm]                  & [\rcm]      & \\
\midrule
\ch{CH3OH}  & 4397 & 2.27 & 4413 (calc.) & $-16$ & $\nu'_{\mathrm{as}}(\mathrm{CH_3}) + \delta_{\mathrm{s}}(\mathrm{CH_3})$ \\
            & 4273 & 2.34 & 4241 (exp.)  & $+32$ & $\delta(\mathrm{COH}) + \nu_{\mathrm{s}}(\mathrm{CH_3})$$^c$ \\
            & 4116 & 2.43 & 4113 (calc.) & $+3$  & $\rho'(\mathrm{CH_3}) + \nu'_{\mathrm{as}}(\mathrm{CH_3})$ \\
            & 4024 & 2.48 & 4022 (calc.) & $+2$  & $\nu(\mathrm{CO}) + \nu_{\mathrm{as}}(\mathrm{CH_3})$ \\
\midrule
\ch{CH3OD}  & 4404 & 2.27 & 4398 (exp.)  & $+6$  & $\delta'_{\mathrm{as}}(\mathrm{CH_3}) + \nu_{\mathrm{as}}(\mathrm{CH_3})$ \\
            & 4280 & 2.34 & 4273 (exp.)  & $+7$  & $2\delta(\mathrm{COD}) + \nu(\mathrm{OD})$$^c$ \\
            & 4224 & 2.37 & 4200 (exp.)  & $+24$ & $\nu'_{\mathrm{as}}(\mathrm{CH_3}) + \rho(\mathrm{CH_3})$ \\
            & 4124 & 2.42 & 4120 (calc.) & $+4$  & $\rho(\mathrm{CH_3}) + \nu_{\mathrm{s}}(\mathrm{CH_3})$ \\
            & 4065 & 2.46 & 4054 (exp.)  & $+11$ & $\rho'(\mathrm{CH_3}) + \nu_{\mathrm{s}}(\mathrm{CH_3})$ \\
\midrule
\ch{CD3OH}  & 4447 & 2.25 & 4469 (calc.) & $-22$ & $2\nu_{\mathrm{as}}(\mathrm{CD_3})$ \\
            & 4393 & 2.28 & 4401 (calc.) & $-8$  & $2\nu'_{\mathrm{as}}(\mathrm{CD_3})$ \\
\midrule
\ch{CD3OD}  & 4459 & 2.24 & 4472 (calc.) & $-13$ & $\nu'_{\mathrm{as}}(\mathrm{CD_3}) + \nu_{\mathrm{as}}(\mathrm{CD_3})$ \\
            & 4440 & 2.25 & 4435 (exp.)  & $+5$  & $2\nu_{\mathrm{as}}(\mathrm{CD_3})$ \\
            & 4326 & 2.31 & 4340 (calc.) & $-14$ & $\nu_{\mathrm{s}}(\mathrm{CD_3}) + \nu_{\mathrm{as}}(\mathrm{CD_3})$ \\
            & 4280 & 2.34 & 4275 (calc.) & $+5$  & $\nu'_{\mathrm{as}}(\mathrm{CD_3}) + \nu_{\mathrm{s}}(\mathrm{CD_3})$ \\
\bottomrule
\end{tabular}
\tablefoot{\textnormal{Band positions are given in \rcm, with corresponding wavelengths in \si{\um}. 
$^{a}$ Closest matching value from \citealt{Grabska2017}; (calc.) denotes B2PLYP/SNST anharmonic VPT2 calculations, and (exp.) experimental measurements in dilute CCl$_4$ solution at 298~K.
$^{b}$ $\Delta$ = (\text{this work}) - \citealt{Grabska2017}, in \rcm.
$^{c}$ Assignments involve modes with significant mixing; see \citealt{Grabska2017} for detailed potential energy distribution analysis. The table follows the notation adopted in that work.}}
\end{table*}

\begin{table}[htbp]
\centering
\caption{Overtone and combination bands of solid methanol and its isotopologues at \SI{10}{\kelvin}, comparing experimentally observed band positions with anharmonic calculated values presented in this work.}
\label{tab:overtones_combinations}
\footnotesize
\setlength{\tabcolsep}{4pt}
\renewcommand{\arraystretch}{1.0}
\begin{tabular}{lcccccl}
\toprule
Isotopologue 
& $\nu_{\mathrm{exp}}$ 
& $\lambda_{\mathrm{exp}}$ 
& $\nu_{\mathrm{calc}}$ 
& $\Delta$ 
& $I_{\mathrm{anh}}$ 
& Assignment \\
& [\rcm] 
& [\um] 
& [\rcm] 
& [\rcm] 
& [km mol$^{-1}$] 
& \\
\midrule
\ch{CH3OH} \\
& 2593 & 3.86 & 2615 & $-22$ & 0.18 & $\rho(\mathrm{CH_3}) + \delta_{\mathrm{s}}(\mathrm{CH_3})$ umbrella \\
& 2528 & 3.96 & 2554 & $-26$ & 0.17 & $\delta$(CH$_3$) scissor + [$\nu$(CO) + $\delta$(OH) + $\delta$(CH$_3$)] \\
& 2442 & 4.10 & 2393 & $+49$ & 0.63 & $\delta(\mathrm{OH}) + [\nu(\mathrm{CO})+\delta(\mathrm{OH})+\delta(\mathrm{CH_3})]$ \\
& 2234 & 4.48 & 2227 & $+7$ & 0.02 & $\rho(\mathrm{CH_3}) + [\nu(\mathrm{CO})+\delta(\mathrm{OH})+\delta(\mathrm{CH_3})]$ \\
& 2042 & 4.90 & 2042 & +0 & 1.51 & $2\nu(\mathrm{CO})$ \\
\midrule
\ch{CH3OD} \\
& 2180 & 4.59 & 2190 & $-10$ & 0.15 & $\rho(\mathrm{CH_3}) + \nu(\mathrm{CO})$ \\
& 2034 & 4.92 & 2043 & $-9$ & 1.94 & $2\nu(\mathrm{CO})$ \\
& 1904 & 5.25 & 1887 & $+17$ & 0.15 & $\nu(\mathrm{CO}) + \delta(\mathrm{OD})$ \\
\midrule
\ch{CH2DOH} \\
& 2570 & 3.89 & 2579 & $-9$ & 0.16 & $\delta(\mathrm{OH}) + \rho(\mathrm{CH_2})$ \\
& 2509 & 3.99 & 2480 & $+29$ & 0.63 & $2\delta(\mathrm{OH})$ \\
& 2368 & 4.22 & 2410 & $-42$ & 0.15 & $\nu(\mathrm{CO}) + \omega(\mathrm{CH_2})$ \\
& 2051 & 4.88 & 2061 & $-10$ & 1.17 & $2\nu(\mathrm{CO})$ \\
& 1947 & 5.14 & 1924 & $+23$ & 0.67 & $\nu(\mathrm{CO}) + \delta(\mathrm{CD})$ \\
\midrule
\ch{CHD2OH} \\
& 2592 & 3.86 & 2629 & $-37$ & 0.48 & $\delta(\mathrm{OH}) + \delta_{\mathrm{s}}(\mathrm{CHD_2})$ umbrella \\
& 1970 & 5.08 & 1990 & $-20$ & 0.22 & $\nu(\mathrm{CO}) + \rho_{\mathrm{as}}(\mathrm{CD_2})$ \\
\midrule
\ch{CD3OH} \\
& 2139 & 4.68 & 2141 & $-2$ & 1.82 & $\delta(\mathrm{OH}) + \rho(\mathrm{CD_3})$ \\
& 2009 & 4.98 & 2016 & $-7$ & 0.48 & $\rho_{\mathrm{as}}(\mathrm{CD_3}) + \delta_{\mathrm{s}}(\mathrm{CD_3})$ umbrella \\
& 1958 & 5.11 & 1960 & $-2$ & 0.97 & $2\nu(\mathrm{CO})$ \\
\midrule
\ch{CD3OD} \\
& 2156 & 4.64 & 2151 & +5 & 1.52 & $\delta$(CD$_3$) scissor + $\delta_{as}$(CD$_3$) \\
& 2012 & 4.97 & 2017 & $-5$ & 0.37 & $\nu(\mathrm{CO}) + \delta(\mathrm{OD})$ \\
& 1936 & 5.17 & 1950 & $-14$ & 0.79 & $2\nu(\mathrm{CO})$ \\
\bottomrule
\end{tabular}
\tablefoot{Experimental band positions are listed in \rcm\ and \si{\um}. For each observed feature, the dominant contributing transition from our anharmonic calculations is reported, together with its calculated frequency ($\nu_{\mathrm{calc}}$), absolute deviation from experiment ($\Delta = \nu_{\mathrm{exp}} - \nu_{\mathrm{calc}}$), and anharmonic infrared intensity ($I_{\mathrm{anh}}$), as well as the corresponding mode assignment. Because this spectral region is strongly anharmonic and heavily mixed, additional weaker overtone and combination contributions are likely present but are not discussed further in this work.}
\end{table}

\clearpage

\section{Lattice modes of deuterated methanol ices at 120~K}
\label{app:lattice_modes}

\begin{figure}[H]
    \centering
    \includegraphics[scale=0.38]{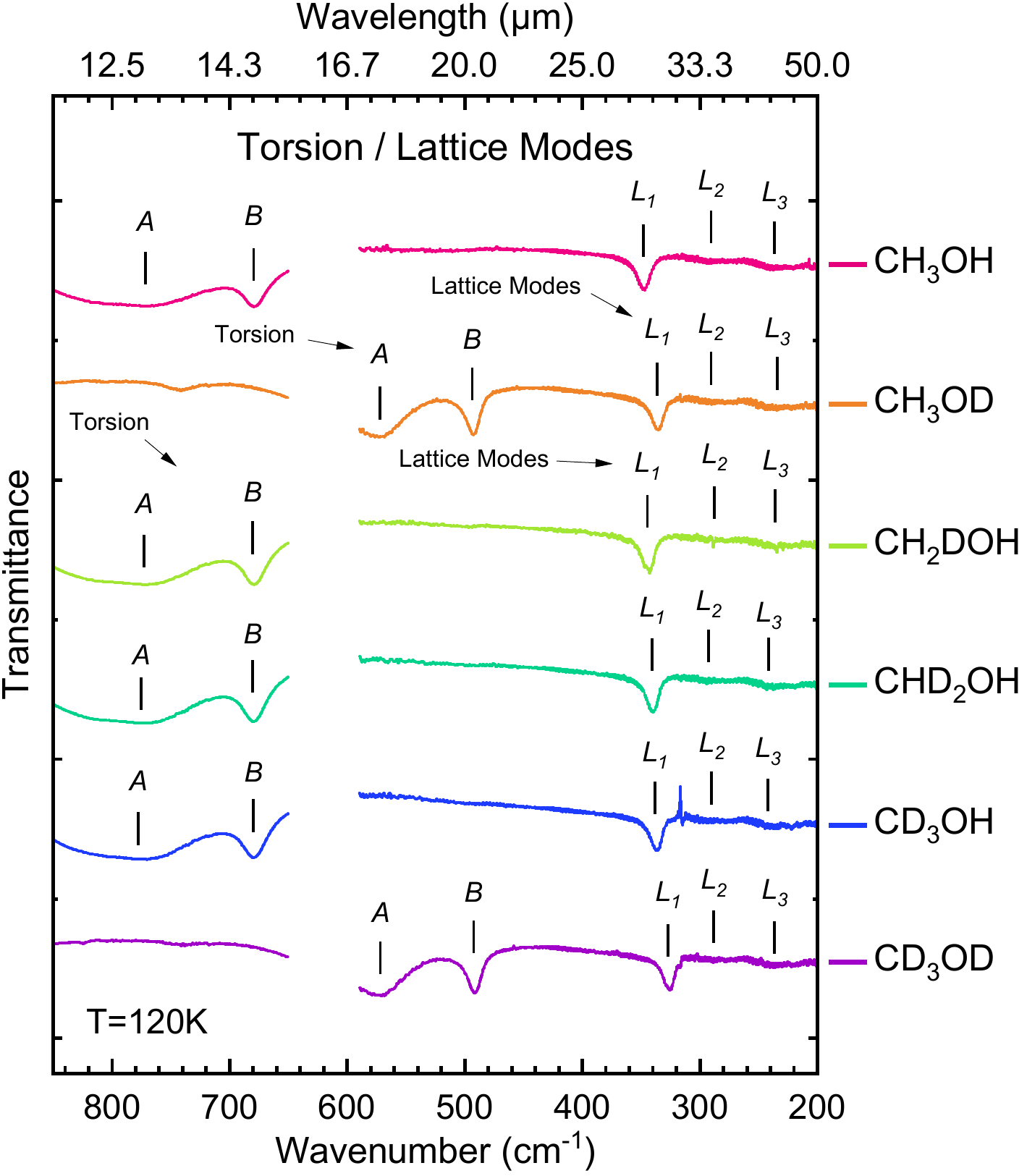}
     \caption{Transmission spectra of crystalline methanol ice (\ch{CH3OH}) and its isotopologues at 120~K in the far-infrared.  The two torsional components as well as the three strongest lattice modes are marked. The spectral region between \SIrange{650}{590}{\rcm} is not shown due to noise introduced by the beamsplitter. Spectra are vertically offset for clarity.}
    \label{fig:mode9_Lattice}
\end{figure}

The low-frequency infrared spectra recorded after annealing to \SI{120}{\kelvin} are
dominated by two classes of motions: (i) the internal methyl
torsion ($\nu_{12}$) and (ii) collective lattice modes of the crystalline ice. In
crystalline \ch{CH3OH}, the torsional mode splits into two components, labelled \textit{A}
and \textit{B} in Fig.~\ref{fig:mode9_Lattice}, appearing at \SI{771}{\rcm} and
\SI{680}{\rcm}, respectively. Substitution of the hydroxyl hydrogen by deuterium,
as in \ch{CH3OD} and \ch{CD3OD}, shifts both components to lower wavenumbers, to
\SI{572}{\rcm} and \SI{493}{\rcm}, consistent with the increased reduced mass
associated with the O--D rotor. In contrast, isotopic substitution confined to
the methyl group (\ch{CH2DOH}, \ch{CHD2OH}, and \ch{CD3OH}) leaves the \textit{A} component close to the \ch{CH3OH} value (\SI{771}{\rcm}), with a modest blue--shift to
\SIrange{772}{778}{\rcm} across the methyl-deuterated isotopologues, while the \textit{B} component remains at \SI{680}{\rcm}. At lower frequencies ($<$~\SI{400}{\rcm}), three lattice bands are observed for all isotopologues. These consist of a higher-frequency feature near \SI{350}{\rcm}
(\textit{L$_1$}) and two lower-frequency bands near \SI{290}{\rcm} (\textit{L$_2$}) and \SI{240}{\rcm} (\textit{L$_3$}). Isotopic substitution leads only to modest red--shifts, typically not exceeding \SI{10}{\rcm}, reflecting the collective nature of these vibrations. The positions of the torsional and lattice features are indicated in Fig.~\ref{fig:mode9_Lattice}, with corresponding frequencies summarised in Table~\ref{tab:mode9_Lattice}.

\begin{table}[H]
\raggedright
\caption{Frequencies of the two torsional components (\textit{A}, \textit{B}) and the three strongest lattice modes (\textit{L}$_1$–\textit{L}$_3$) of crystalline methanol and its isotopologues at \SI{120}{\kelvin}. }
\label{tab:mode9_Lattice}
\footnotesize
\setlength{\tabcolsep}{4pt}
\renewcommand{\arraystretch}{1.05}
\begin{tabular*}{\columnwidth}{@{\extracolsep{\fill}}lcccccc}
\toprule
Mode & \ch{CH3OH} & \ch{CH3OD} & \ch{CH2DOH} & \ch{CHD2OH} & \ch{CD3OH} & \ch{CD3OD} \\
\midrule
\textit{A}     & 771 & 572 & 772 & 776 & 778 & 572 \\
\textit{B}     & 680 & 493 & 680 & 680 & 680 & 492 \\
\textit{L}$_1$ & 348 & 336 & 344 & 341 & 338 & 327 \\
\textit{L}$_2$ & 293 & 290 & 287 & 293 & 291 & 288 \\
\textit{L}$_3$ & 237 & 234 & 236 & 243 & 242 & 237 \\
\bottomrule
\end{tabular*}
\tablefoot{Band positions are given in \rcm.}
\end{table}

\FloatBarrier

\end{document}